\documentclass[letterpaper]{article} 
\usepackage{aaai24}  
\usepackage{times}  
\usepackage{helvet}  
\usepackage{courier}  
\usepackage[hyphens]{url}  
\usepackage{graphicx} 
\usepackage{natbib}  
\usepackage{caption} 
\usepackage{subfig}
\usepackage{algorithm}
\usepackage{algorithmic}
\usepackage{bbding}
\usepackage{booktabs}
\usepackage{enumerate}
\usepackage{amsthm}
\usepackage{multirow}
\usepackage{color}
\usepackage{framed}
\usepackage{verbatim}
\usepackage{thmtools} 
\usepackage{thm-restate}
\usepackage{epsfig}
\usepackage{amsmath}
\usepackage{amssymb}
\usepackage{graphics}

\newtheorem{definition}{Definition}
\newtheorem{lemma}{Lemma}
\usepackage[capitalize]{cleveref}
\crefname{section}{Sec.}{Secs.}
\Crefname{section}{Section}{Sections}
\Crefname{table}{Table}{Tables}
\crefname{table}{Tab.}{Tabs.}

\usepackage{newfloat}
\usepackage{listings}
\DeclareCaptionStyle{ruled}{labelfont=normalfont,labelsep=colon,strut=off} 
\floatstyle{ruled}
\newfloat{listing}{tb}{lst}{}
\floatname{listing}{Listing}
\title{Task-Agnostic Privacy-Preserving Representation Learning for Federated Learning Against Attribute Inference Attacks}

\author{
    Caridad Arroyo Arevalo\textsuperscript{\rm 1},
    Sayedeh Leila Noorbakhsh\textsuperscript{\rm 1},
    Yun Dong\textsuperscript{\rm 2},
    Yuan Hong\textsuperscript{\rm 3},
    Binghui Wang\textsuperscript{\rm 1}
}
\affiliations{

    \textsuperscript{\rm 1}Illinois Institute or Technology, 
    \textsuperscript{\rm 2}Benedictine University, 
    \textsuperscript{\rm 3}University of Connecticut \\
    carroyoarevalo@hawk.iit.edu, snoorbakhsh@hawk.iit.edu, ydong@ben.edu, yuan.hong@uconn.edu, bwang70@iit.edu
}

\usepackage{bibentry}

\begin{document}

\maketitle

\begin{abstract}

Federated learning (FL)  has been widely studied recently due to its property to collaboratively train data from different devices without sharing the raw  data. Nevertheless, recent studies show that an adversary can still be possible to infer private information about devices' data, e.g., sensitive attributes such as income, race, and sexual orientation. 
To mitigate the attribute inference attacks, various 
existing privacy-preserving FL methods can be adopted/adapted. 
However, all these existing methods have key limitations: they need to know the FL task in advance, or have intolerable computational overheads or utility losses, or do not have provable privacy guarantees. 

We  address these issues and design a {task-agnostic} privacy-preserving presentation learning method for FL ({\bf TAPPFL}) against attribute inference attacks. TAPPFL is formulated via information theory. Specifically, 
TAPPFL has two mutual information goals, where one goal learns task-agnostic data representations that contain the least information about the private attribute in each device's data, and the other goal ensures the learnt data representations include as much information as possible about the device
data to maintain FL utility. 
We also derive privacy guarantees of TAPPFL against worst-case attribute inference attacks, as well as the inherent tradeoff between utility preservation and privacy protection.
Extensive results on multiple datasets and applications validate the effectiveness of TAPPFL to protect data privacy, maintain the FL utility, and be efficient as well. 
Experimental results also show that TAPPFL outperforms the existing defenses\footnote{Source code and full version: \url{https://github.com/TAPPFL}}. 

\end{abstract}
\section{Introduction}
\label{sec:intro}

The emerging collaborative  data  analysis  using federated  learning  (FL)~\cite{mcmahan2017communication} 
aims to address the data insufficiency problem, and has a great potential to protect data privacy as well. 
In FL, the participating devices train their data locally, and only share the trained models,
instead of the raw data, with a center server (e.g., cloud).   
The server updates its global model by aggregating the received device models, and broadcasts the updated global model to all participating devices such that all devices \emph{indirectly} use all data from other devices.
FL has been deployed by many companies such as Google~\cite{GoogleFL}, Microsoft~\cite{MSFL}, IBM~\cite{IBMFL}, and Alibaba~\cite{AliFL}, and applied in various privacy-sensitive applications, including on-device item ranking~\cite{mcmahan2017communication}, content suggestions for on-device keyboards~\cite{bonawitz2019towards}, next word prediction~\cite{li2020federated}, health monitoring~\cite{rieke2020future}, and medical imaging~\cite{kaissis2020secure}.
However, recent works show,  though only sharing device models, it is still possible for an adversary (e.g., an honest-but-curious server) to perform the \textit{attribute inference attack}
~\cite{jia2017attriinfer,aono2017privacy,ganju2018property,melis2019exploiting,dang2021revealing,wainakh2022user}---i.e., inferring the private/sensitive information  
(e.g.,  a person's gender, race, sexual orientation, income) of device's data. 
Hence, 
designing privacy-preserving FL mechanisms to defend against the attribute inference attack is important and necessary.

To mitigate the issue, various existing privacy-preserving 
FL methods can be adopted/adapted, including \textit{multi-party computation (MPC)}~\cite{danner2015fully,mohassel2017secureml,bonawitz2017practical,melis2019exploiting}, \textit{adversarial training (AT)}~\cite{madras2018learning,liu2019privacy,li2019deepobfuscator,oh2017adversarial,kim2019training}, \textit{model compression (MC)}~\cite{zhu2019deep}, and \textit{differential privacy (DP) }~\cite{pathak2010multiparty,shokri2015privacy,hamm2016learning,mcmahan2018learning,geyer2017differentially}. 
However, these existing methods have key limitations, thus narrowing their applicability 
(see Table \ref{tb:dpfl}). 
Specifically, MPC and AT methods know the FL task 
in advance. However, this cannot be achieved in real-world unsupervised learning applications.    
MPC methods also have intolerable computational overheads and AT methods do not have provable privacy guarantees. 
MC and DP methods are task-agnostic, but both of them result in high utility losses (see Figure \ref{fig:defense_res}). 

In this paper, we aim to design a 
privacy-preserving FL mechanism against attribute inference attacks  (termed {\bf TAPPFL}) that is \emph{task-agnostic}, \emph{efficient}, \emph{accurate}, and have \emph{privacy guarantees} as well. Our main idea is to learn federated privacy-preserving representations based on information theory. 
Specifically, we formulate TAPPFL via two mutual information (MI) goals, where one MI goal learns 
low-dimensional representations for device data that contain the least information about the
private attribute in each device's data---thus protecting attribute privacy, and the other MI goal ensures the learnt representations include as much information as possible about the training data---thus maintaining FL utility.  
Our TAPPFL is task-agnostic as our formulation does not need to know the FL task.  
However, the true MI values are challenging to compute,
due to that they deal with high-dimensional random variables and require to compute an intractable posterior distribution. 
Inspired by the MI neural estimators~\cite{belghazi2018mutual,chen2016infogan,cheng2020club}, we recast calculating intractable exact MI values into deriving tractable (variational) MI bounds, where 
each variational bound is associated with a posterior distribution that 
can be parameterized via a neural network. Hence, estimating the true MI values reduces to training the parameterized neural networks. 
We further propose an alternative learning algorithm to train these neural networks and learn task-agnostic privacy-preserving representations for device data. We also derive  privacy
guarantees of TAPPFL against worst-case attribute inference attacks, as well as 
the inherent tradeoff between utility preservation and attribute privacy protection. 
Finally, we evaluate TAPPFL on multiple datasets and applications. 
Our results validate the learnt devices' data representations can be used to achieve high utility and maintain attribute privacy as well. 

Our key contributions are highlighted as follows:
\begin{itemize}
\vspace{-1mm}
\item {\bf Algorithm:} We propose the first practical privacy-preserving FL mechanism (TAPPFL), i.e., task-agnostic, efficient, and accurate, against attribute inference attacks.

\vspace{-1mm}
    \item {\bf Theory:} TAPPFL has privacy guarantees and shows an inherent tradeoff between utility and privacy. 

   \vspace{-1mm}
    \item {\bf Evaluation:} 
    TAPPFL is effective against attribute inference attacks and shows  advantages over the baselines.  \vspace{-2mm}
\end{itemize}

\begin{table}[!t]\renewcommand{\arraystretch}{0.75}
\footnotesize
\center
\caption{Comparisons of the PPFL methods.}
\addtolength{\tabcolsep}{-3pt}
\vspace{-2mm}
\label{tb:dpfl}
\begin{tabular}{|c|c|c|c|c|}
\hline
 {\bf Methods} & {\bf Task-Agnostic} & {\bf Efficient} & {\bf Provable} & {\bf Accurate}  \\  \hline 
{\bf MPC} &  &  & \checkmark &\checkmark  \\ \hline
{\bf AT} &  & \checkmark&  &\checkmark \\ \hline
{\bf MC}  & \checkmark & \checkmark & & \\ \hline
{\bf  DP}  &  \checkmark& \checkmark & \checkmark &   \\ \hline\hline
{\bf TAPPFL} & \checkmark  &  \checkmark & \checkmark &  \checkmark \\ \hline
\end{tabular}
\vspace{-6mm}
\end{table}    
\section{Related Work}
\noindent {\bf Privacy-preserving FL against inference attacks.}
\textit{Secure multi-party computation}~\cite{danner2015fully,mohassel2017secureml,bonawitz2017practical,melis2019exploiting},   \textit{adversarial training}~\cite{oh2017adversarial,wu2018towards,madras2018learning,pittaluga2019learning,liu2019privacy,kim2019training}, \textit{model compression}~\cite{zhu2019deep}, and \textit{differential privacy (DP)}~\cite{pathak2010multiparty,shokri2015privacy,hamm2016learning,mcmahan2018learning,geyer2017differentially,wei2020federated} are the four typical privacy-preserving FL methods. 
For example, \citet{bonawitz2017practical} design a secure multi-party aggregation for FL, where devices are required to encrypt their local models before uploading them to the server. 
However, it incurs an intolerable computational overhead and may need to know the specific FL task in advance. 
Adversarial training methods
are inspired by GAN~\cite{goodfellow2014generative}. 
These methods adopt adversarial learning to 
learn obfuscated features from the training data so that their privacy information cannot be inferred from a learnt model. However, these methods also need to know the FL task and lack of formal privacy guarantees. 
\citet{zhu2019deep} apply gradient compression/sparsification to defend against privacy leakage from shared local models. However, to achieve a desirable privacy protection, such approaches require high compression rates, leading to intolerable utility losses. In addition, it does not have formal privacy guarantees. 
\citet{shokri2015privacy} propose a collaborative learning method where the sparse vector is adopted to achieve DP.
However, DP methods have high utility losses. 

\vspace{+0.01in}
\noindent {\bf Mutual information for fair representation learning.} 
\citet{moyer2018invariant,song2019learning} leverage MI to perform fair representation learning. 
The goal of fair representation learning is to encode the input data into a representation that aims to mitigate bias, e.g., demographic disparity, 
towards a private group/attribute. 
For instance, when achieving demographic parity, the predicted label should be independent of the private attribute. 
For instance, 
\cite{moyer2018invariant} proposed to use the information bottleneck objective~\cite{alemi2017deep} 
and train a variational auto-encoder~\cite{kingma2019introduction} to censor a private attribute by minimizing a variational bound on the MI between the learnt representations and the private attribute. 

{Besides the difference of the studied problem (fairness vs privacy-protection), other key differences are: these fair methods need to know the learning task  
in advance, and only estimate the MI for low-dimensional variables.  In contrast, our method is task-agnostic and designs a novel MI estimator to 
handle the challenging high-dimensional variables. Further, 
these methods {do not} have {theoretical} fairness guarantees, while ours has formal privacy guarantees.}

\vspace{+0.01in}
\noindent {\bf Mutual information for private representation learning.}
The most closely related to our work is DPFE~\cite{osia2018deep}, which also formulates learning private representations via MI objectives. However, there are major differences: i) DPFE knows the primary task, i.e., \emph{task-specific}, while ours is task-agnostic; 
ii) DPFE does not have formal privacy guarantees, while ours do have; 
iii) DPFE estimates MI by making assumptions on distributions of the random variable, e.g., Gaussian. 
In contrast, our method derives MI {bounds} on random variables whose distributions can be \emph{arbitrary}, 
and then trains networks to approximate the true MI. 
 
\vspace{+0.01in}
\noindent {\bf Mutual information  estimation.}
Accurately estimating MI between any random variables is challenging~\cite{belghazi2018mutual}.
Recent methods~\cite{alemi2017deep,belghazi2018mutual,oord2018representation,poole2019variational,hjelm2019learning,cheng2020club,wang2021privacy} propose to first derive 
(upper or lower) MI
bounds by introducing auxiliary variational  distributions and then train parameterized neural networks to estimate variational distributions and approximate true MI.  
For instance, MINE~\cite{belghazi2018mutual}
views MI as a KL divergence between the joint and marginal distributions, converts it into the dual representation, and derives a lower MI bound. 
\citet{cheng2020club} design 
a Contrastive Log-ratio Upper Bound of MI, which connects MI with contrastive learning~\cite{oord2018representation}, and estimates MI as the difference of conditional probabilities between positive and negative sample pairs.

\section{Background and Problem Definition}
\label{sec:background}

{\bf Federated learning (FL).} 
The FL paradigm enables a server to coordinate the training of multiple local devices through multiple rounds of global communications, without sharing the local data. 
Suppose there are $M$ devices $\mathcal{C} = \{C_i\}_{i=1}^{M}$ and a server $S$ participating in  FL. Each device $C_i$ has data samples
$\mathbf{x}^i$ from a distribution $\mathcal{D}^i$ over the sample space $\mathcal{X}^i$. 
In each round $t$, each device $C_{i}$ first downloads the previous round's global model (e.g., $\Theta_{t-1}$) from the server, and  then updates its local model (e.g.,  $\Theta_{t}^i$) using the local data $\{\mathbf{x}^i \}$ and global model $\Theta_{t-1}$. 
The server $S$ then randomly collects a set of (e.g., $K$) current local  models in devices (e.g., $\mathcal{C}_K$) and updates the global model for the next round using an aggregation algorithm. For example, when using the most common  FedAvg~\cite{McMahan17}, the server updates the global model as 
$\Theta_{t}\gets  \sum_{i \in \mathcal{C}_K} \frac{n_{i}}{\sum_{i\in \mathcal{C}_k} n_i} \Theta_{t}^{i}$, where $n_{i}$ is the size of the training data of device $C_{i}$.

\noindent {\bf Threat model and problem definition.}
We assume each device $C_i$'s data has its \emph{private attribute} and denote it as $u^i$.
Each device $C_i$ aims to learn a feature extractor $f_{\Theta^i}: \mathcal{X}^i \rightarrow \mathcal{R}^i$, parameterized by $\Theta^i$, that maps data samples from input space $\mathcal{X}^i$ to the latent representation space $\mathcal{R}^i$; and we denote the learnt 
representation for a sample $\mathbf{x}^i$ as $\mathbf{r}^i = f_{\Theta^i}(\mathbf{x}^i)$. 
The learnt representations can be used for  downstream tasks, e.g.,  next-word-prediction on smart phones~\cite{li2020federated}. 
We assume the server $S$ is honest-but-curious and it has access to the learnt representations $\{{\bf r}^i\}$ of device data. 
The server's purpose is to infer any {private attribute} ${\bf u}^i$ through the 
representations without tampering the FL training process. 
Our goal is to learn the feature extractor $f_{\Theta^i}$ per device such that it  protects the private attribute $u^i$ from being inferred, as well as preserving the primary FL task utility. 
For a general purpose, we assume the primary FL task is unknown (i.e., task-agnostic)  to the defender (i.e., who learns the feature extractor). 
W.l.o.g, the protected attribute is different from the primary task label. 

\section{Design of TAPPFL}
\label{sec:TAPPFL}

In this section, we will design our task-agnostic privacy-preserving FL (TAPPFL) method against attribute inference attacks. 
Our TAPPFL is inspired by information theory. We show that our TAPPFL has provable privacy guarantees, as well as inherent utility-privacy tradeoffs. 

\vspace{-2mm}
\subsection{Formulating TAPPFL via MI Objectives}
For ease of description, we choose a device $C_i$ and 
show how to learn its privacy-preserving feature extractor $f_{\Theta^i}$. 
Our goal is to transform the data $\mathbf{x}^i \sim \mathcal{D}^i$ into a representation $\mathbf{r}^i = f_{\Theta^i}(\mathbf{x}^i) $ that satisfies the below two goals: 

\begin{itemize}

\item \textbf{Goal 1: Privacy protection.} 
$\mathbf{r}^i$ contains as less information as possible about 
the private attribute $u^i$. Ideally, when $\mathbf{r}^i$ does not include information about 
$u^i$, it is impossible for the server to infer $u^i$ from $\mathbf{r}^i$. 

\item \textbf{Goal 2: Utility preservation.} 
$\mathbf{r}^i$ includes as much information about 
the raw data $\mathbf{x}^i$ as possible. Ideally, when $\mathbf{r}^i$ retains the most information about $\mathbf{x}^i$, the model trained on $\mathbf{r}^i$ will have the same performance as the model trained on the raw $\mathbf{x}^i$, thus preserving utility. 
Note that this goal does not know the FL task, thus task-agnostic. 
\vspace{-1mm}
\end{itemize}
We propose to formalize the above two goals via MI objectives. 
In information theory, MI is a measure of shared information between two random variables, and offers a metric to quantify the ``amount of information" obtained about one random variable by observing the other
random variable.
Formally, we quantify the privacy protection and utility reservation goals using two MI 
objectives as follows:
{
\begin{align}
    & 
    \textbf{Achieving Goal 1: }\quad \min\nolimits_{\Theta^i} I(\mathbf{r}^i; u^i); \quad 
    \notag \\
    & 
    \textbf{Achieving Goal 2: } \quad \max\nolimits_{\Theta^i} I(\mathbf{x}^i; \mathbf{r}^i | u^i). 
    \label{eqn:pprl}
\end{align}
}%
where 
$I(\mathbf{r}^i; u^i)$ is the MI between $\mathbf{r}^i$ and $u^i$, and we minimize such MI to maximally reduce the correlation between $\mathbf{r}^i$ and $u^i$.
$ I(\mathbf{x}^i; \mathbf{r}^i | u^i) $ is the MI between $\mathbf{x}^i$ and $\mathbf{r}^i$ given $u^i$. We maximize such MI to keep the raw information in $\mathbf{x}^i$ as much as possible in $\mathbf{r}^i$ and remove the information that $\mathbf{x}^i$ contains about the private $u^i$ to leak into ${\bf r}^i$.

\vspace{-2mm}
\subsection{Estimating MI via tractable variational bounds}
The key challenge of solving the above two MI objectives is that 
calculating an MI between two arbitrary random variables is likely to be 
infeasible \cite{peng2018variational}. 
To address it, we are inspired by the existing MI neural estimation methods~\cite{alemi2017deep,belghazi2018mutual,oord2018representation,poole2019variational,hjelm2019learning,cheng2020club}, which convert the intractable exact MI calculations to the tractable variational MI bounds. 
Specifically, we first obtain a MI upper bound for privacy protection and a MI lower bound for utility preserving via introducing two auxiliary posterior distributions, respectively.  
Then, we parameterize each auxiliary distribution with a neural network, and approximate the true posteriors by minimizing the upper bound  and  maximizing the  lower bound through training the involved neural networks. 

\vspace{+0.01in}
\noindent {\bf Minimizing upper bound MI for privacy protection.} 
We propose to adapt the variational upper bound CLUB proposed   in~\cite{cheng2020club} to bound $I(\mathbf{r}^i;u^i)$. Specifically, 
\begin{small}
\begin{align}
        & I(\mathbf{r}^i;u^i) \leq  I_{vCLUB}(\mathbf{r}^i;u^i)  \notag \\
        = & \mathbb{E}_{p(\mathbf{r}^i, u^i)} [\log q_{\Psi^i}(u^i|\mathbf{r}^i) ] - \mathbb{E}_{ p(\mathbf{r}^i) p(u^i)} [\log q_{\Psi^i}(u^i|\mathbf{r}^i) ],
\end{align}
\end{small}%
where 
$q_{\Psi^i}(u^i | \mathbf{r}^i)$ is an auxiliary posterior distribution of $p(u^i | \mathbf{r}^i)$ needing to satisfy the  condition: 
$$KL (p(\mathbf{r}^i, u^i) ||  q_{\Psi^i} (\mathbf{r}^i, u^i)) \leq KL (p(\mathbf{r}^i)p(u^i) || q_{\Psi^i} (\mathbf{r}^i, u^i)),$$
where $KL[q(\cdot)||p(\cdot)]$ is the Kullback-Leibler divergence between two distributions $q(\cdot)$ and $p(\cdot)$ and is nonnegative. 
To achieve this, we thus 
need to minimize:
\begin{small}
\begin{align}
\small
    & \min_{\Psi^i}  KL (p(\mathbf{r}^i, u^i) ||  q_{\Psi^i} (\mathbf{r}^i, u^i)) \notag \\
    = &\min_{\Psi^i}  KL (p(u^i | \mathbf{r}^i) ||  q_{\Psi^i} (u^i | \mathbf{r}^i)) \notag \\
    =& \min_{\Psi^i}  \mathbb{E}_{    p(\mathbf{r}^i, u^i)} [\log p(u^i| \mathbf{r}^i)] -  \mathbb{E}_{    p(\mathbf{r}^i, u^i)} [\log q_{\Psi^i}(u^i| \mathbf{r}^i))] \notag \\
\Leftrightarrow  &  \max_{\Psi^i}    \mathbb{E}_{ p(\mathbf{r}^i, u^i)} [\log q_{\Psi^i}(u^i| \mathbf{r}^i)],  \label{eqn:KL_conv} 
\end{align}
\end{small}%
where we use that the first term 
$\mathbb{E}_{p(\mathbf{r}^i, u^i)} [\log p(u^i| \mathbf{r}^i)]$ 
in the second-to-last Equation 
is irrelevant to $\Psi^i$.   

Finally, our {\bf Goal 1} for privacy protection is reformulated as solving the below 
min-max objective function: 
\begin{small}
\begin{align}
\small
        & \min \limits_{\Theta^i} I(\mathbf{r}^i;u^i) 
        = \min \limits_{\Theta^i} \min \limits_{\Psi^i}  I_{vCLUB}(\mathbf{r}^i;u^i)  \notag \\
        & \Longleftrightarrow \min \limits_{\Theta^i} \max \limits_{\Psi^i} 
        \mathbb{E}_{    p(\mathbf{r}^i, u^i)} [\log q_{\Psi^i}(u^i| \mathbf{r}^i)].
        \label{eqn:prot_adv}
\end{align}
\end{small}%
\emph{Remark.} Equation (\ref{eqn:prot_adv}) can be interpreted as an \emph{adversarial game} between: (1) an adversary $q_{\Psi^i}$ (i.e., attribute inference classifier) who aims to infer the private attribute $u^i$ from $\mathbf{r}^i$; and (2) a defender (i.e., the feature extractor $f_{\Theta^i}$) who aims to protect $u^i$  from being inferred from $\mathbf{r}^i$.

\vspace{+0.01in}
\noindent {\bf Maximizing lower bound MI for utility preservation.}
We derive the lower bound of the MI 
$ I(\mathbf{x}^i; \mathbf{r}^i | u^i)$ as follows: 
\begin{small}
\begin{align}
    & I({\bf x}^i ; {\bf r}^i | u^i ) = 
    H({\bf x}^i | u^i) - H({\bf x}^i | {\bf r}^i, u^i) \notag \\ 
    & = H({\bf x}^i | u^i) + \mathbb{E}_{p({\bf x}^i , {\bf r}^i, u)} [\log p({\bf x}^i | {\bf r}^i, u^i))] \notag \\
    & = H({\bf x}^i | u^i) + \mathbb{E}_{p({\bf x}^i , {\bf r}^i, u)} [\log q_{\Omega^i} ({\bf x}^i | {\bf r}^i, u^i))] \notag \\ 
    & \qquad + \mathbb{E}_{p({\bf x}^i , {\bf r}^i, u)} [KL(p(\cdot|{\bf r}^i, u^i) || q_{\Omega^i} (\cdot | {\bf r}^i, u^i))]  \notag \\
    & \geq H({\bf x}^i | u^i) + \mathbb{E}_{p({\bf x}^i , {\bf r}^i, u)} [\log q_{\Omega^i} ({\bf x}^i | {\bf r}^i, u^i))],  
\end{align}
\end{small}%
where $q_{\Omega^i}$ is an arbitrary auxiliary posterior distribution that aims to maintain the information in $\mathbf{x}^i$, and $H(\mathbf{x}^i | u^i)$ is the condition entropy 
that is constant. 
Hence, our {\bf Goal 2} can be rewritten as the following max-max objective function:
{
\small
\begin{align}
    & \max_{\Theta^i} I(\mathbf{x}^i; \mathbf{r}^i | u^i) 
    \Longleftrightarrow
    \max_{\Theta^i} \max_{\Omega^i} \mathop{\mathbb{E}}_{p(\mathbf{x}^i, \mathbf{r}^i, u^i)} \left[\log {q_{\Omega^i}}[(\mathbf{x}^i | \mathbf{r}^i, u^i)\right]. 
    \label{eqn:max_jsd}
\end{align} 
}%
\emph{Remark.} 
Equation (\ref{eqn:max_jsd}) can be interpreted as a \emph{cooperative game} between the feature extractor $f_{\Theta^i}$ and $q_{\Omega^i}$ who aim to preserve the utility collaboratively. 

\vspace{+0.01in}
\noindent {\bf Objective function of TAPPFL.} By combining Equations (\ref{eqn:prot_adv}) and (\ref{eqn:max_jsd}) and considering all devices, 
 our final objective function of learning the task-agnostic privacy-preserving representations in FL is as follows:
 {
\small
\begin{align} 
    & \sum_{C_i \in \mathcal{C}} \max_{\Theta^i}\Big(\lambda_i  \min_{\Psi^i} - \mathbb{E}_{p(u^i,\mathbf{x}^i)}\left[\log q_{\Psi^i}(u^i | f_{\Theta^i}(\mathbf{x}^i)) \right] \notag \\ 
    & \, {+} (1-\lambda_i) \max_{\Omega^i} \mathbb{E}_{p(\mathbf{x}^i, u^i)} \left[\log {q_{\Omega^i}}[(\mathbf{x}^i | f_{\Theta^i}(\mathbf{x}^i), u^i)\right] \Big) 
    \label{eq:final-obj}, 
\end{align}
}%
where $\lambda_i \in [0,1]$ achieves a tradeoff between privacy and utility for device $C_i$. That is, a larger $\lambda_i$ indicates a stronger attribute privacy protection for $C_i$'s data, while a smaller $\lambda_i$ indicates a better utility preservation for $C_i$'s data. 

\begin{figure}[!t]
\vspace{-2mm}
    \centering
    \subfloat[TAPPFL scheme]{\includegraphics[scale=0.065]{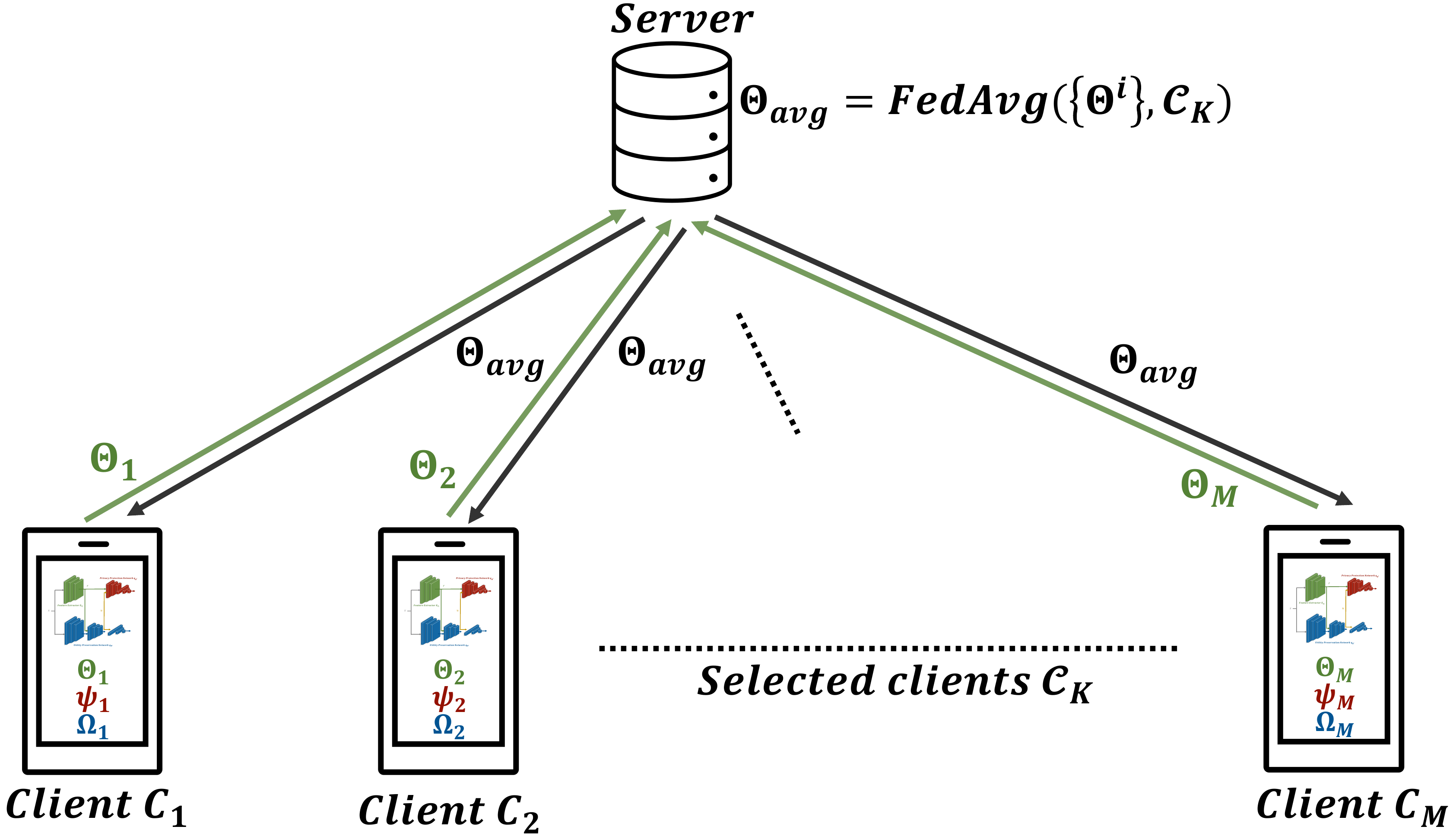} \label{fig:aa}}
	\subfloat[TAPPFL device training]{\centering\includegraphics[scale=0.07]{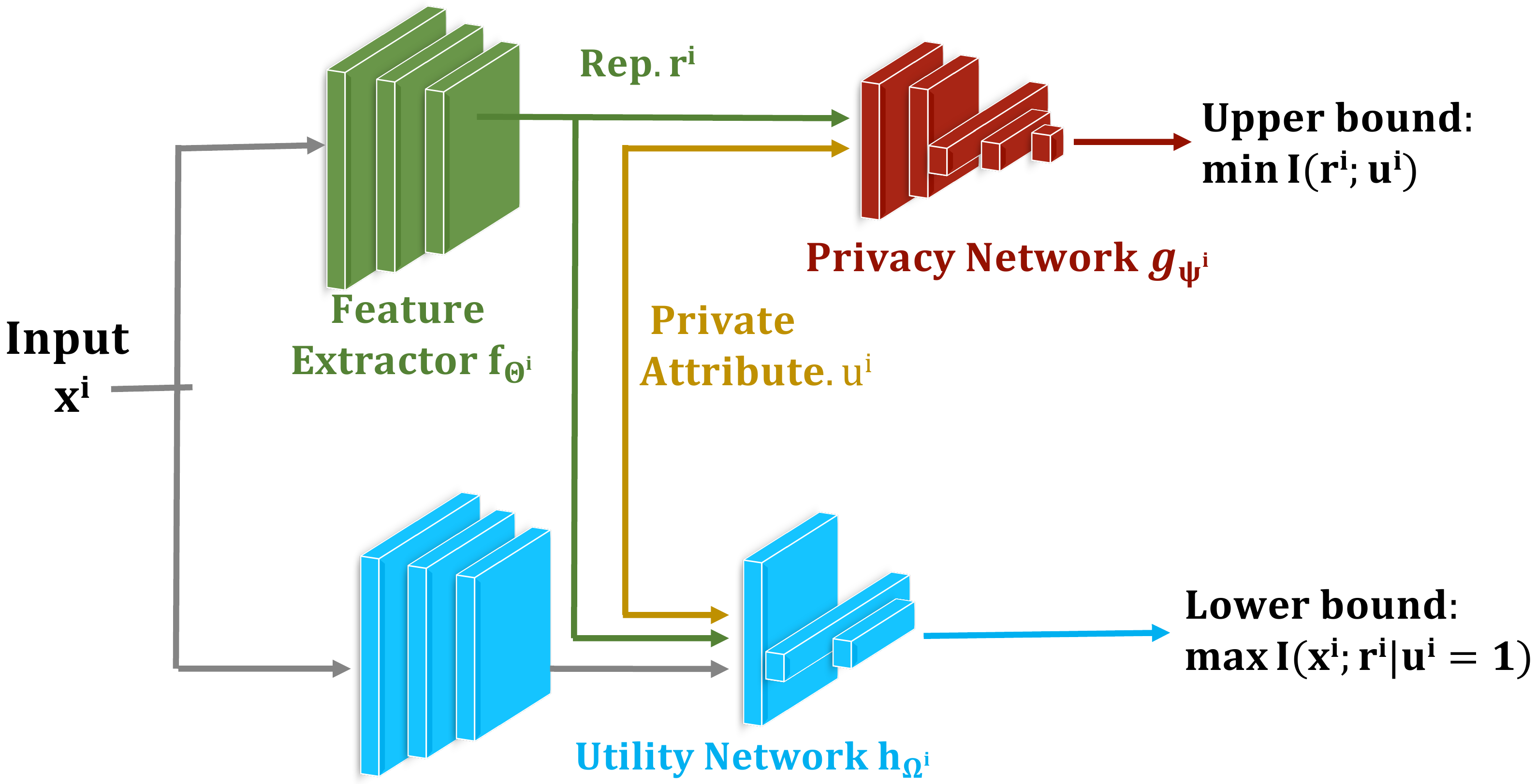} \label{fig:bb}}  
	\vspace{-3mm}
	\caption{\small (a) TAPPFL scheme; (b) Training device training.}
	\vspace{-7mm}
	\label{fig:diagram}
\end{figure}

\vspace{-2mm}
\subsection{Implementation in Practice}
Directly calculating the exact expectation in Equation (\ref{eq:final-obj}) is challenging.  
In practice, Equation (\ref{eq:final-obj}) can be solved 
via training three parameterized neural networks,  i.e., the feature extractor $f_{\Theta^i}$, the privacy-protection network $g_{\Psi^i}$ associated with the auxiliary distribution $q_{\Psi^i}$, and the utility-preservation network  $h_{\Omega^i}$ associated with the auxiliary distribution $q_{\Omega^i}$, using sampled data from each device $C_i$. 
Specifically, in each device $C_i$, we first collect a set of samples $\{{\bf x}^i_j\}$ and the associated private attributes $\{u^i_j\}$ from $\mathcal{D}^i$. Note that, as our TAPPFL is task-agnostic, we do not know the sample labels for the FL task. 
With it, we can then approximate the expectation terms in Equation (\ref{eq:final-obj}). 

Specifically, we approximate the first expectation term as 
\begin{small}
\begin{align*}
& \mathbb{E}_{p(u^i, \mathbf{x}^i)}\log q_{\Psi^i}(u^i|f_{\Theta^i}(\mathbf{x}^i)) \approx
 -\sum\nolimits_{j} CE(u^i_j,g_{\Psi^i}(f_{\Theta^i}(\mathbf{x}^i_j))),
 \end{align*}
 \end{small}%
 where $CE(\cdot)$ means the cross-entropy error function.

 Moreover, note that the data ${\bf x}$ and the representation  ${\bf r}$ are rather high-dimensional. To address it, we propose to use the \textit{Jensen-Shannon} divergence (JSD)  \cite{hjelm2019learning} for high-dimensional MI estimation and approximate the second expectation term associated with $q_{\Omega^i}$ as below:  
\begin{small}
\begin{align*}
& \mathbb{E}_{p(\mathbf{x}^i, u^i)}\log q_{\Omega^i}(\mathbf{x}^i|f_{\Theta^i}(\mathbf{x}_i),u^i)  =  I^{(JSD)}_{\Theta^i,\Omega^i} (\mathbf{x}^i; f_{\Theta^i}(\mathbf{x}^i),u^i) \notag  \\  
&= \mathbb{E}_{(\mathbf{x}^i,u^i)}[-\textrm{sp}(-h_{\Omega^i}(\mathbf{x}^i,f_{\Theta^i}(\mathbf{x}^i),u^i)] \notag \\ & \quad - \mathbb{E}_{(\mathbf{x}^i,u^i, \bar{\bf x}^i)}[\textrm{sp}(h_{\Omega^i}(\bar{\bf x}^i,f_\Theta(\mathbf{x}^i),u^i)],
\end{align*}
\end{small}%
where $\textrm{sp}(z) = \log(1+\exp(z))$ is the softplus function, $\bar{\bf x}^i$ is an independent 
sample from the same distribution as ${\bf x}^i$, and the expectation can be replaced by samples $\{{\bf x}_j^i, \bar{\bf x}_j^i, u_j^i \}$.

Figure \ref{fig:diagram} illustrates our TAPPFL.
It needs to simultaneously train three neural networks, i.e., the feature extractor $f_{\Theta^i}$, the privacy-protection network $g_{\Psi^i}$, and the utility-preservation network $h_{\Omega^i}$, in each device $C_i$.  
In particular, 
 the server first initializes a global model $\Theta_{0}$ and broadcasts $\Theta_{0}$ to all devices; and the devices initializes $\{\Psi^i_{0}\}$ and $\{\Omega^i_{0}\}$ locally. 
 Then the training procedure  involves two iterative steps. 
 For example, in the $t$-th round: 
In Step I, each device updates ${\Theta^i_{t}}$ using the received $\Theta_{t-1}$ from the server, and locally updates $\Psi^i_{t}$ and $\Omega^i_{t}$ using $\Psi^i_{t-1}$ and $\Omega^i_{t-1}$  
based on its training data; and the devices send the updated  $\{\Theta^i_{t}\}$ to the server. 
In Step II, the server selects  a set of $\{\Theta^i_{t}\}$ and updates the global model $\Theta_{t}$ by aggregating these models via, 
e.g., Fedvg~\cite{McMahan17}, and broadcasts $\Theta_{t}$ to all devices.
We repeat the two steps alternately until convergence or reaching the maximum number of iterations.
Algorithm~\ref{alg:tappfl} 
in Appendix details the TAPPFL training process. 

\begin{table*}[!t]
\footnotesize
\centering
   \caption{Test accuracy (Test Acc.) vs. attribute inference accuracy (Infer. Acc.) on the considered three datasets.}
      \vspace{-2mm}
\addtolength{\tabcolsep}{-3pt}
\centering
    \begin{tabular}{l|c|c|c||c|c|c||c|c|c||c|c|c}
      \hline
           \multicolumn{4}{c||}{\bf CIFAR10} & \multicolumn{3}{c||}{\bf Loans} & \multicolumn{3}{c||}{\bf Adult income} & \multicolumn{3}{c}{\bf Adult income} \\
          \hline
          \multicolumn{4}{c||}{\bf Private attribute: Animal or not (binary)} & \multicolumn{3}{c||}{\bf Race (binary)} & \multicolumn{3}{c||}{\bf Gender (binary)} &  \multicolumn{3}{c}{\bf Marital status (7 values)} \\ \hline           
     \textbf{$\lambda$} & {Test Acc} & {Infer. Acc} &{{Gap}} & {Test Acc} & {Infer. Acc} &{{Gap}} & {Test Acc} & {Infer. Acc} &{{Gap}} & {Test Acc} & {Infer. Acc} &{{Gap}} \\
     \hline
     \hline
     0 &  0.89 &  0.74 & 0.24  & 0.98 &   0.74  & 0.24 & 0.825 &   0.700 & 0.20 &  0.825 &   0.375 & 0.232\\
     0.25 & 0.88 & 0.64 & 0.14 & 0.93 &   0.72 & 0.22 &  0.750 &   0.550 & 0.05 &  0.800 &   0.275 & 0.112\\
     0.5 & 0.76 & 0.60 & 0.10 & 0.88 &   0.70 & 0.20 &  0.750 &   0.550 & 0.05 &  0.800 &   0.250 & 0.107\\
     0.75 & 0.67 &  0.56 & 0.06 &  0.84 &   0.63& 0.13 & 0.725 &   0.550 & 0.05 &  0.725 &   0.243 & 0.043\\
     1 & 0.58 & 0.52 & 0.02 & 0.82 &   0.57 & 0.07 & 0.700  &   0.525 & 0.025 & 0.700 &   0.175 & 0.032\\
     \hline 
   \end{tabular} 
   \label{tab:allresults}
   \vspace{-5mm}
\end{table*}

\vspace{-2mm}
\section{Theoretical Results}
\label{sec:theory}

{\bf Inherent utility-privacy tradeoff.}
We consider the attribute has a binary value and the primary FL task is binary classification. We will leave it as a future work to generalize our results to multi-value attributes and multi-class classification.  

Let $\mathcal{A}$ be the set of all binary attribute inference classifiers, i.e., $\mathcal{A}=\{A: {\bf r}^i \in \mathcal{R}^i \rightarrow \{0,1\}, \forall C_i\}$. 
Let $\mathcal{D}^i$ be a joint distribution over the input ${\bf x}^i$, sensitive attribute $u^i$, and label ${\bf y}^i$ for device $C_i$. 
W.l.o.g, we assume the representation space is bounded, i.e., $\max_{C_i \in \mathcal{C}} \max_{{\bf r}^i \in \mathcal{R}^i} \|{\bf r}^i \| \leq R$.
Moreover, we denote the binary task classifier as $c: {\bf r}^i \rightarrow \{0,1\}$, which predicts data labels based on the learnt representation.
We further define the \emph{advantage} of the {worst-case} adversary with respect to 
the joint distribution 
$\mathcal{D}^i$ as below:
\begin{small}
\begin{align}
\small
    \label{eqn:adv}
    & \textrm{Adv}_{\mathcal{D}^i} (\mathcal{A}) = \max_{A \in \mathcal{A}} | \textrm{Pr}_{\mathcal{D}^i}(A({\bf r}^i) =a | u^i=a) \notag \\ 
    & \qquad - \textrm{Pr}_{\mathcal{D}^i} (A({\bf r}^i) =a | u^i=1-a) |, \, \forall a=\{0,1\}. 
\end{align}
\end{small}%
If $\textrm{Adv}_{\mathcal{D}^i}(\mathcal{A}) =1$, this means an adversary can \emph{completely} infer the privacy attribute through the learnt representations. 
In contrast, $\textrm{Adv}_{\mathcal{D}^i}(\mathcal{A})=0$ means an adversary obtains a \emph{random guessing} inference performance. Our goal is thus to learn the representations such that $\textrm{Adv}_{\mathcal{D}^i}(\mathcal{A})$ is small. 
The proofs are in the full version: \url{https://github.com/TAPPFL}.  

\begin{restatable}[]{theorem}{uptradeoff}
\label{thm:uptradeoff}
\vspace{-1mm}
Let ${\bf r}^i$ be the representation with a bounded norm $R$ 
(i.e., $\max_{{\bf r}^i \in \mathcal{R}^i} \|{\bf r}^i \| \leq R$) 
learnt by the feature extractor $f_{\Theta_i}$ for device $C_i$'s data ${\bf x}^i$, 
and $\mathcal{A}$ be the set of all binary attribute inference classifiers.
Assume the task classifier $c$ is $C_L$-Lipschitz, i.e., $\|c\|_L \leq C_L$. Then, each $C_i$'s classification error $\textrm{err}_i$ can be bounded as below:  
\begin{align}
\label{eqn:tradeoff}
    \textrm{err}_i 
    \geq \Delta_{{\bf y}^i|u^i} - 2R \cdot C_L \cdot \textrm{Adv}_{\mathcal{D}^i}(\mathcal{A}),
\end{align}
where $\Delta_{y^i|u^i} = |\textrm{Pr}_{\mathcal{D}^i}(y^i=1|u^i=0) - \textrm{Pr}_{\mathcal{D}^i}(y^i=1|u^i=1)|$ is a device-dependent constant. 
\vspace{-1mm}
\end{restatable}
\noindent \emph{Remark.} 
Theorem~\ref{thm:uptradeoff} states that, for a device-dependent constant $\Delta_{{\bf y}^i|u^i}$, any primary task classifier using   representations learnt by the feature extractor $f_{\Theta_i}$ has to 
incurs a classification error 
on at least a private attribute---The smaller/larger the advantage $\textrm{Adv}_{\mathcal{D}_i}(\mathcal{A})$ is, the larger/smaller the  lower error bound. Note that the lower bound is {independent} of the adversary, meaning it covers the \emph{worst-case}  adversary. Thus, Equation (\ref{eqn:tradeoff}) reflects an inherent trade-off between  utility preservation and attribute privacy leakage. 

\vspace{+0.01in}
\noindent {\bf Privacy guarantees against attribute inference.} 
The attribute inference accuracy incurred by the
worst-case adversary is bounded in the following theorem:

\begin{restatable}[]{theorem}{provprivacy}
\label{thm:provprivacy}
\vspace{-1mm}
Let ${\Theta_*^i}$ (resp. ${\bf r}^i_*$ ) be the learnt optimal feature extractor parameters (resp. optimal representations)  by  Equation (\ref{eq:final-obj}) for device $C_i$'s data. 
Define $H^i_* = H(u^i | {\bf r}^i_*)$. Then, 
for any attribute inference adversary $\mathcal{A}= \{A: {\bf r}^i \rightarrow u^i \}$, 
$\textrm{Pr}({A}({\bf r}^i_*) = u^i) \leq 1 - \frac{H^i_*}{2 \log_2 ({6}/{H_i^*})}$. 
\vspace{-1mm}
\end{restatable}

\noindent \emph{Remark.} Theorem~\ref{thm:provprivacy} shows that when the conditional entropy $H^i_* = H(u^i | {\bf r}^i_*)$ is larger, the attribute inference accuracy induced by any adversary is smaller, i.e., the less attribute privacy is leaked. 
From another perspective, as $H(u^i | {\bf r}^i_*) = H(u^i) - I(u^i;  {\bf r}^i_*)$, achieving the largest $H(u^i | {\bf r}^i_*)$ indicates minimizing the mutual information $I(u^i;  {\bf r}^i_*)$---This is exactly our {\bf Goal 1} aims to achieve.

\section{Experiments}
\label{sec:eval}

\subsection{Experimental setup}

\noindent {\bf Datasets and applications.}
We evaluate our TAPPFL using three datasets from different applications.  CIFAR-10 \cite{Krizhevsky09learningmultiple} is a widespread image dataset. 
The primary task is to predict the label of the image, while the private attribute is a binary attribute indicating if an image belongs to an animal or not.  
For the Loans dataset \cite{10.5555/3157382.3157469}, the primary task is to accurately predict the affordability of the person asking for the loan while protecting their race. 
Finally, for the Adult income dataset \cite{Dua:2019}, predicting whether the income of a person is above \$50,000 or not is the primary task. The private attributes are the gender and the marital status.  More detailed descriptions of these datasets and the training/testing sets can be found in the full version. 

\vspace{+0.01in}
\noindent {\bf Parameter settings.}
We use a total of 100 devices participating in FL training.  By default, the server randomly selects 10\% devices and uses FedAvg~\cite{McMahan17} to aggregate devices' feature extractor parameters in each round. In each device, we train the three parameterized neural networks via the  Stochastic Gradient Descent (SGD) algorithm, where we set the local batch size to be 10 and use 10 local epochs, and the learning rate in SGD is 0.01. A detailed architecture of each neural network can be found in Table \ref{tab:arch} in the full version. 
Before the learning, we first pretrain the 
feature extractor network only to obtain a good initialization, i.e., high utility. 
The number of global rounds is set to be 20. In TAPPFL, for simplicity, we set $\lambda_i = \lambda$ for all devices and all devices share the same private attribute.  
The TAPPFL algorithm is implemented in PyTorch. We use the Chameleon Cloud platform 
offered by the NSF \cite{keahey2020lessons} ({CentOS7-CUDA 11 with Nvidia Rtx 6000}). 

\vspace{+0.01in}
\noindent {\bf Evaluation metrics.} 
We evaluate TAPPFL in terms of 
both utility preservation and privacy protection. 
We use the testing accuracy (i.e., device's feature extractor + utility network on the primary task's test set) to measure utility preservation; and 
attribute inference accuracy (i.e., device's feature extractor + privacy network on the privacy task's test set) to measure the privacy leakage. The larger testing accuracy, the better utility preservation; 
the inference accuracy closer to random guessing, the less attribute privacy leakage.

\begin{figure*}[t]
	\centering
	\subfloat[{CIFAR10: Animal or not}]
	{\centering \includegraphics[scale=0.27]{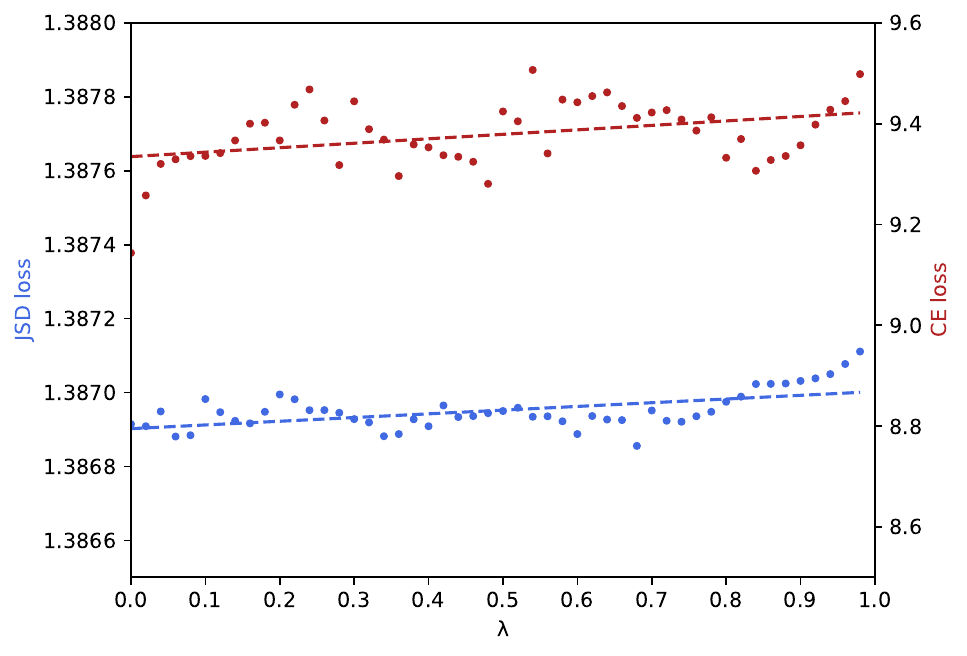}}
	\subfloat[Loans: {Race}]
	{\centering \includegraphics[scale=0.27]{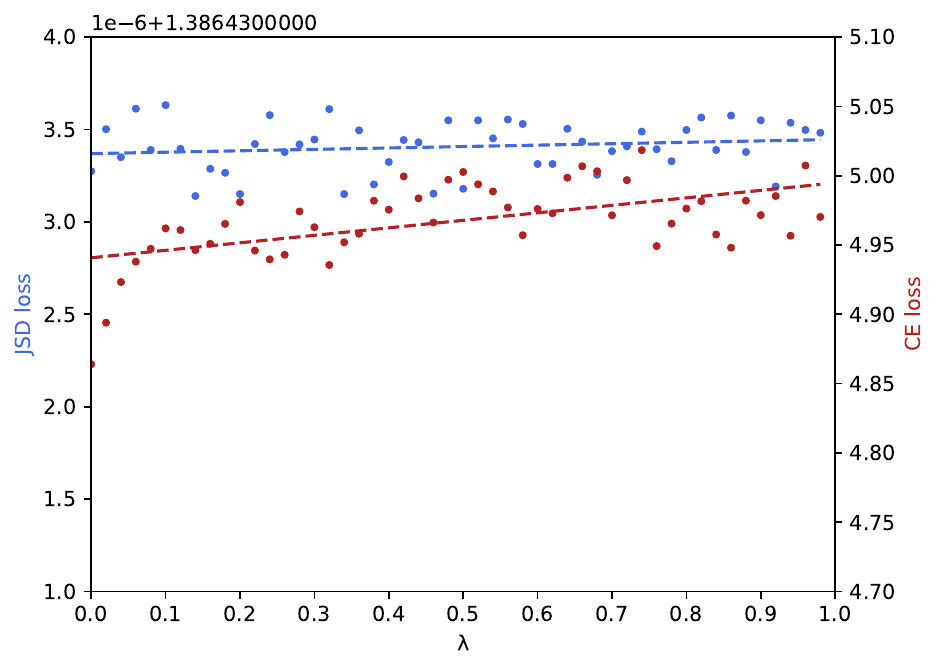}}
	\subfloat[{Adult: Gender}]
	{\centering\includegraphics[scale=0.27]{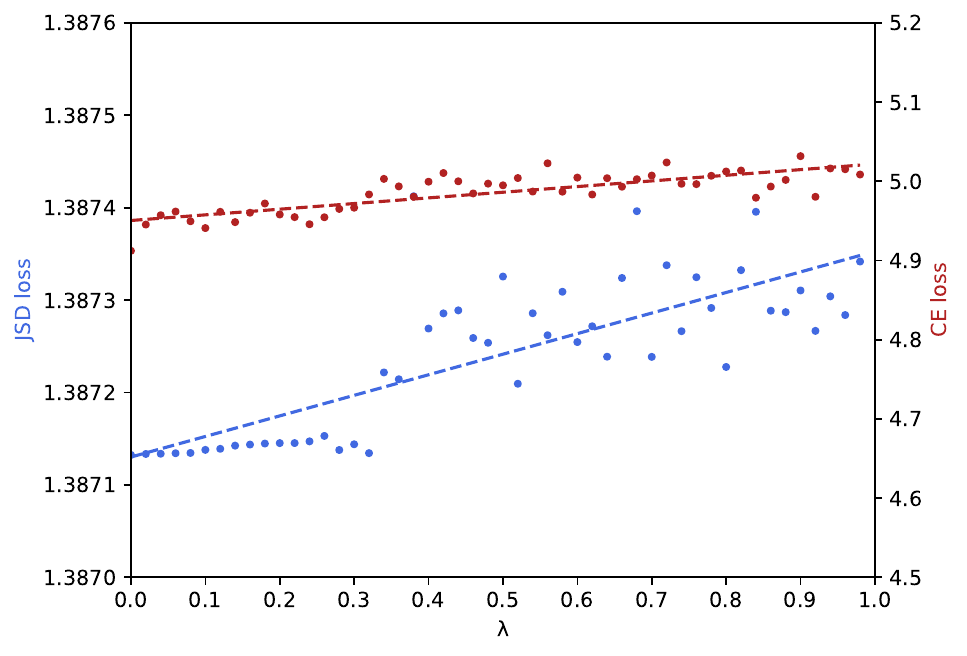}}
	\subfloat[ {Adult: Marital status}]
	{\centering\includegraphics[scale=0.27]{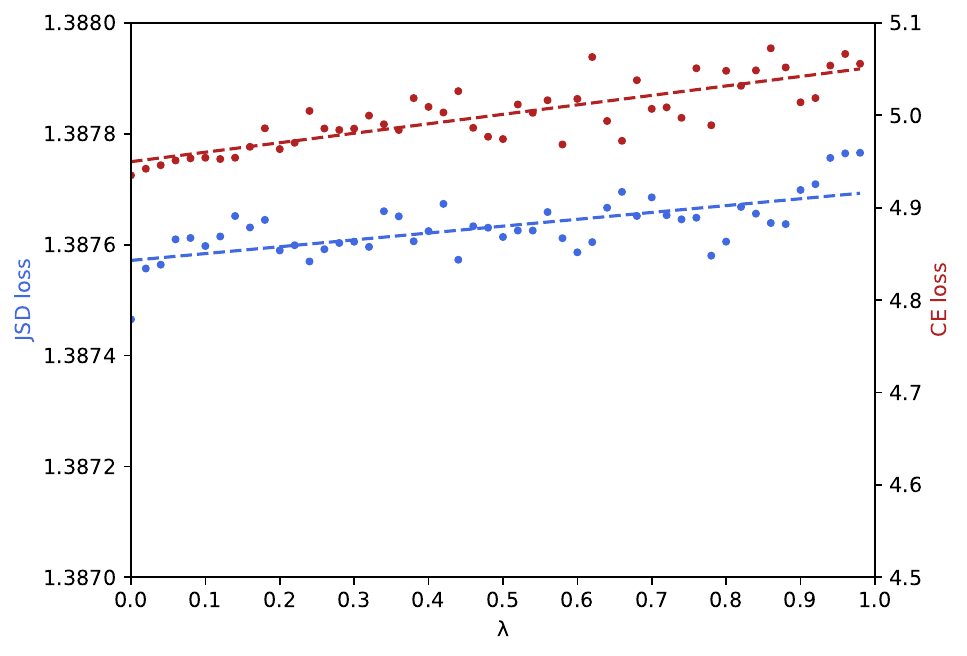}}
	\vspace{-2mm}
	\caption{Mutual information vs. $\lambda$ on a randomly selected client. Note that the CE loss and JSD loss are inversely proportional to the MIs in the two goals. Each point corresponds to a  CE loss or JSD loss at a selected $\lambda$.} 
	\label{fig:plotMIs}
	\vspace{-5mm}
\end{figure*}

\begin{figure*}[!t]
	\centering
	\subfloat[{Adult (G.): Raw input}]
	{\centering\includegraphics[scale=0.20]{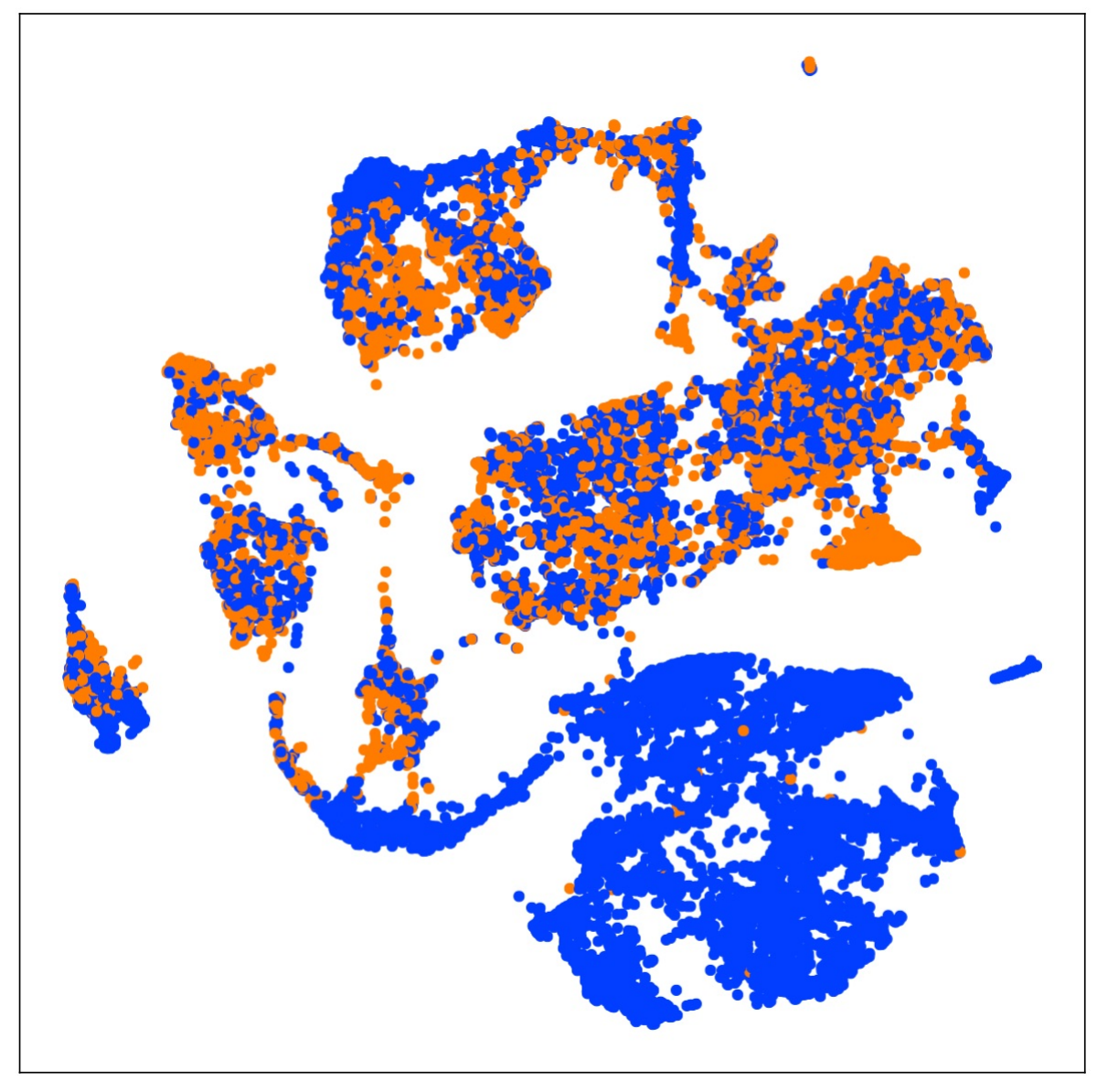}}
	\subfloat[{Adult (G.): Learnt rep.}]
	{\centering \includegraphics[scale=0.20]{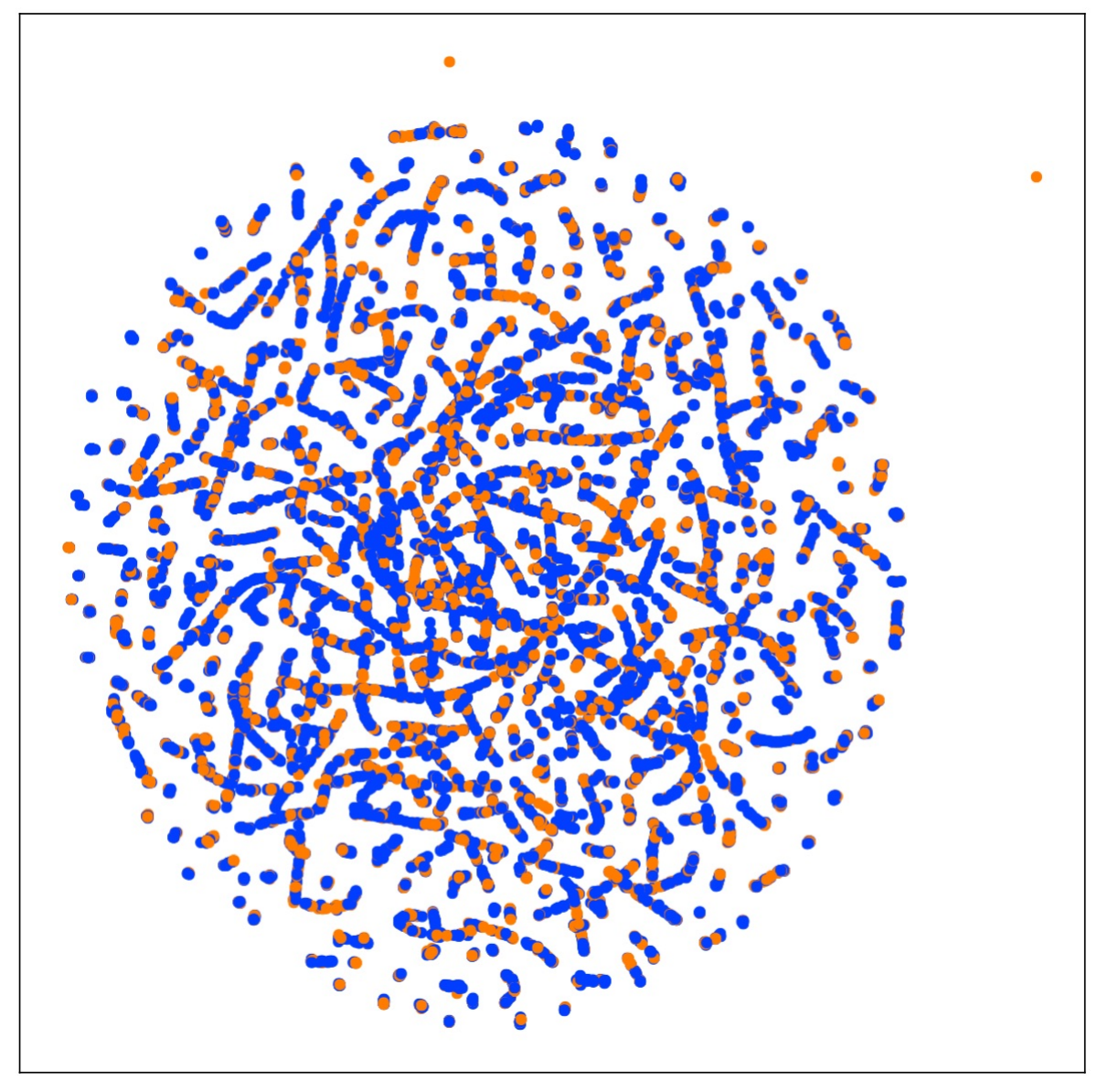}}
	\subfloat[{Adult (M.): Raw input}]
	{\centering\includegraphics[scale=0.20]{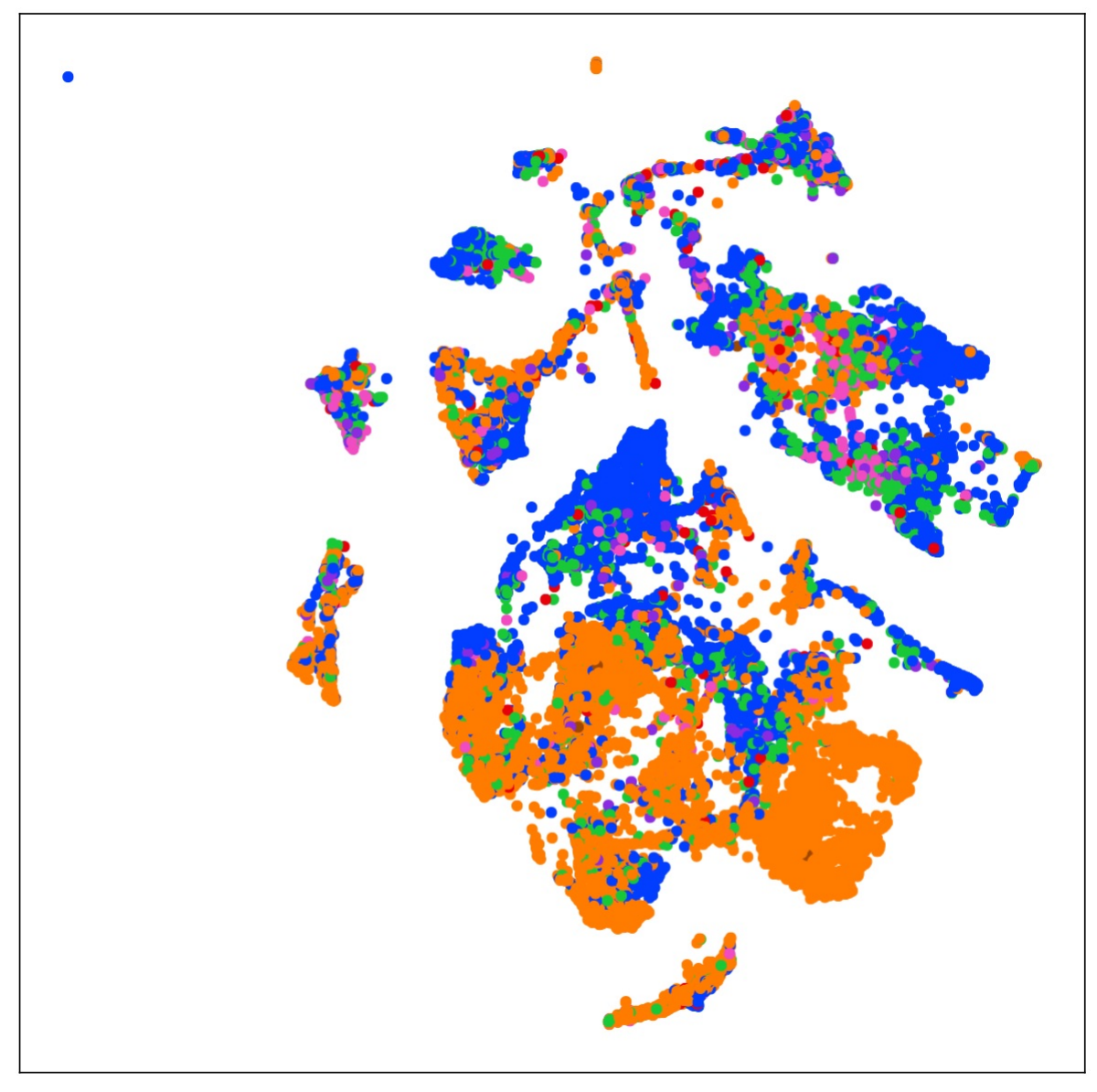}}
	\subfloat[{Adult (M.): Learnt rep.}]
	{\centering \includegraphics[scale=0.20]{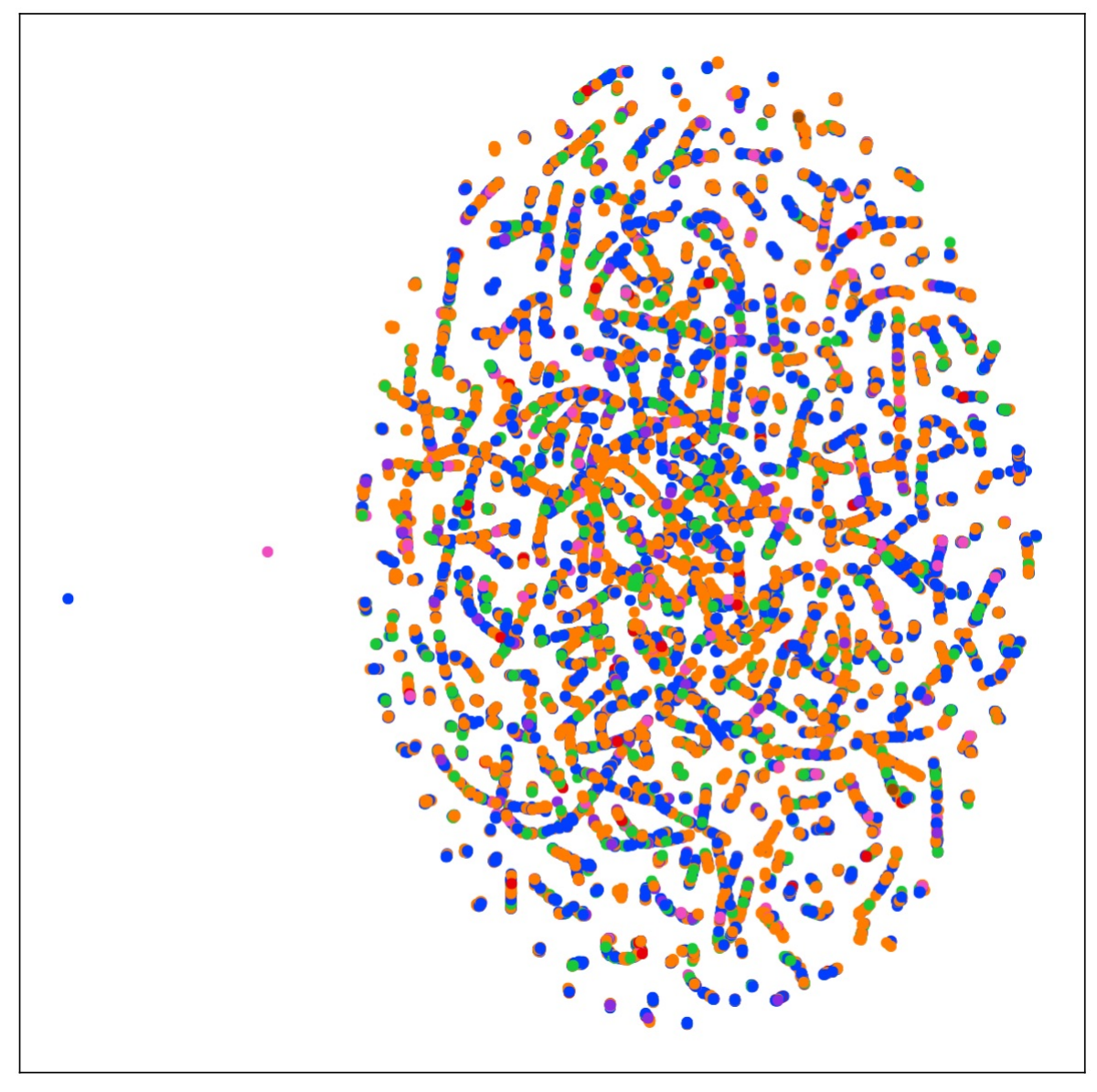}}
	\vspace{-2mm}
	\subfloat[{CIFAR10: Raw input}]
	{\centering\includegraphics[scale=0.20]{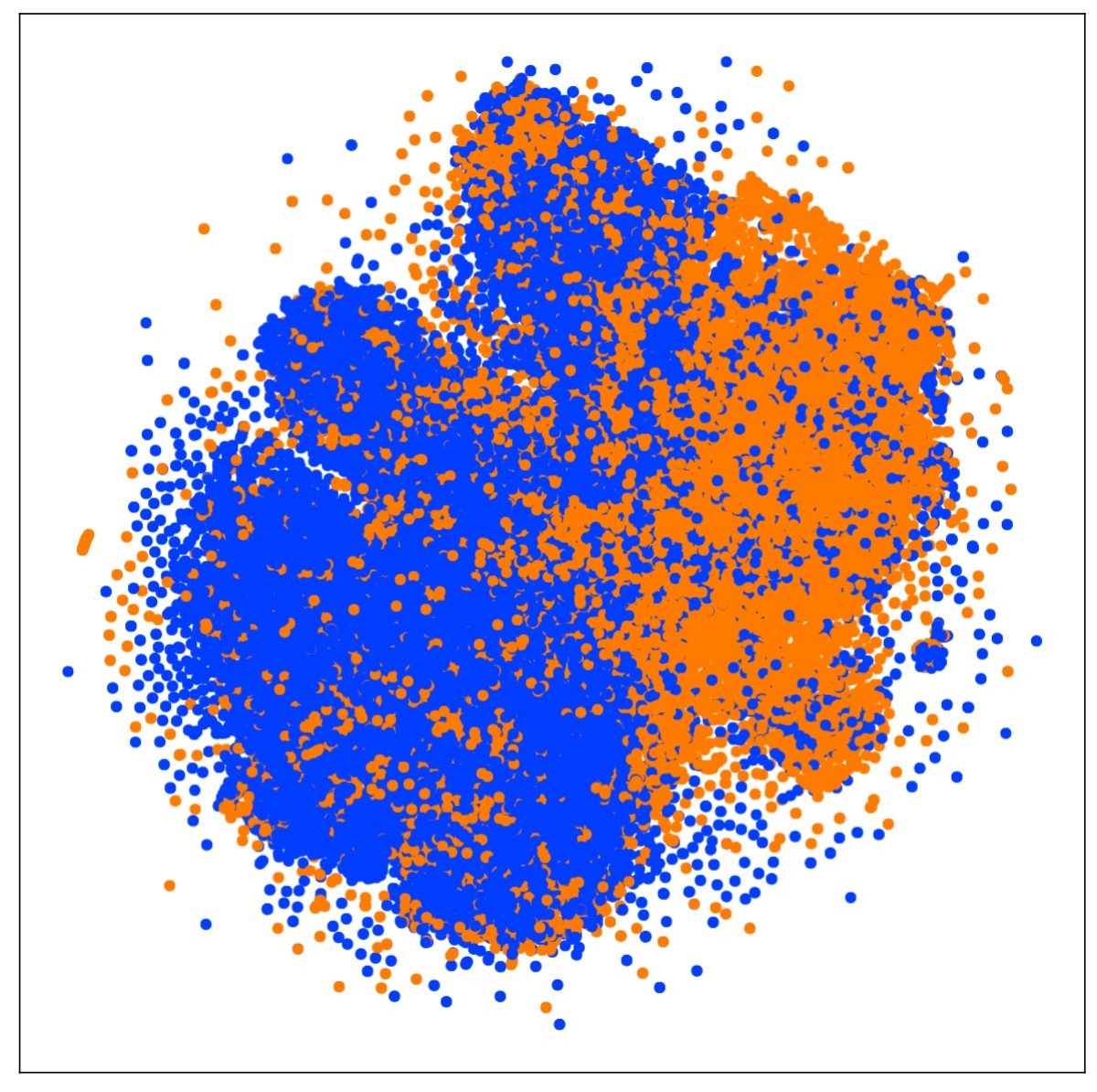}}
	\subfloat[{CIFAR10: Learnt rep.}]
	{\centering \includegraphics[scale=0.20]{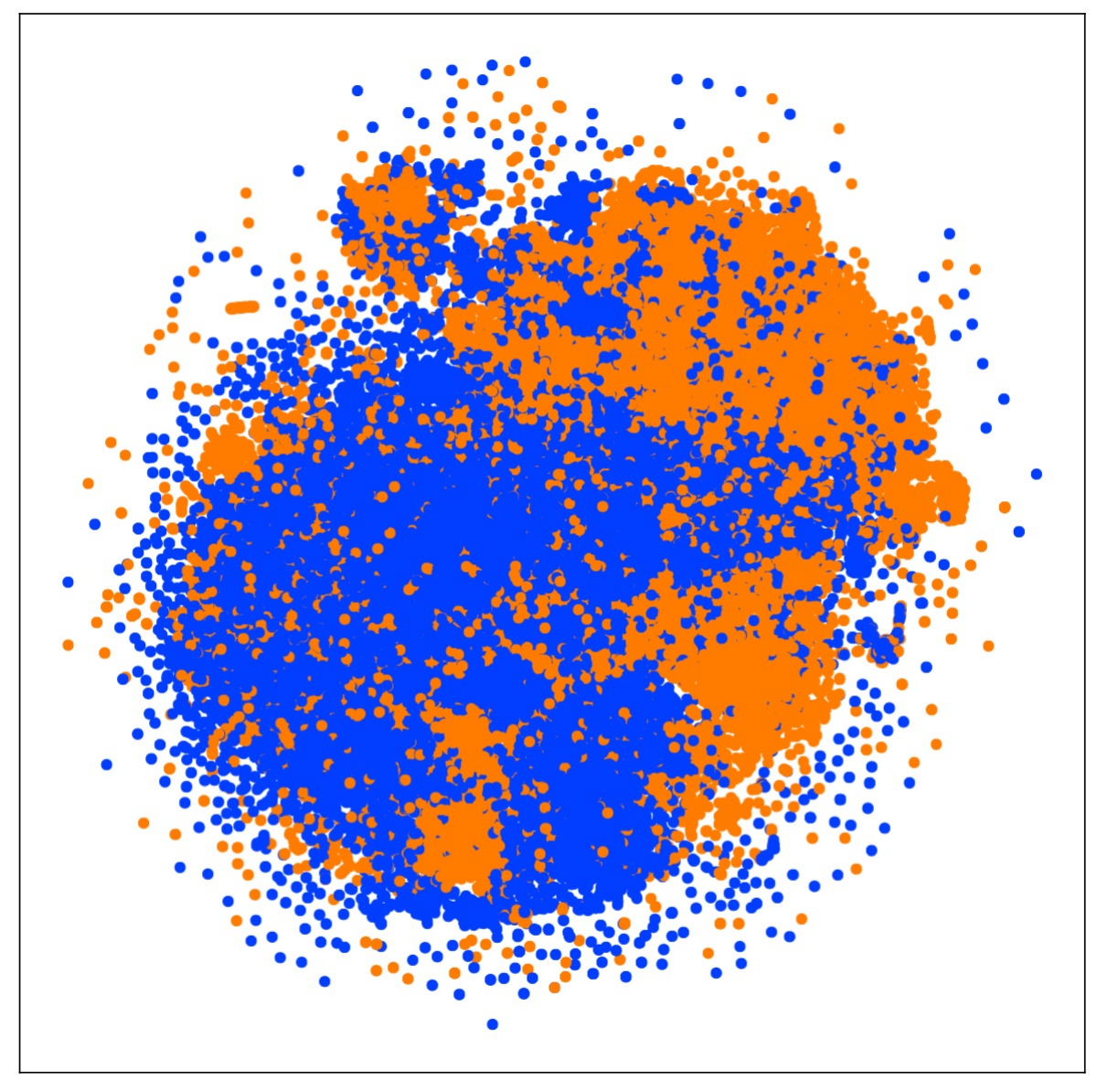}}
	\subfloat[{Loans: Raw input}]
	{\centering\includegraphics[scale=0.20]{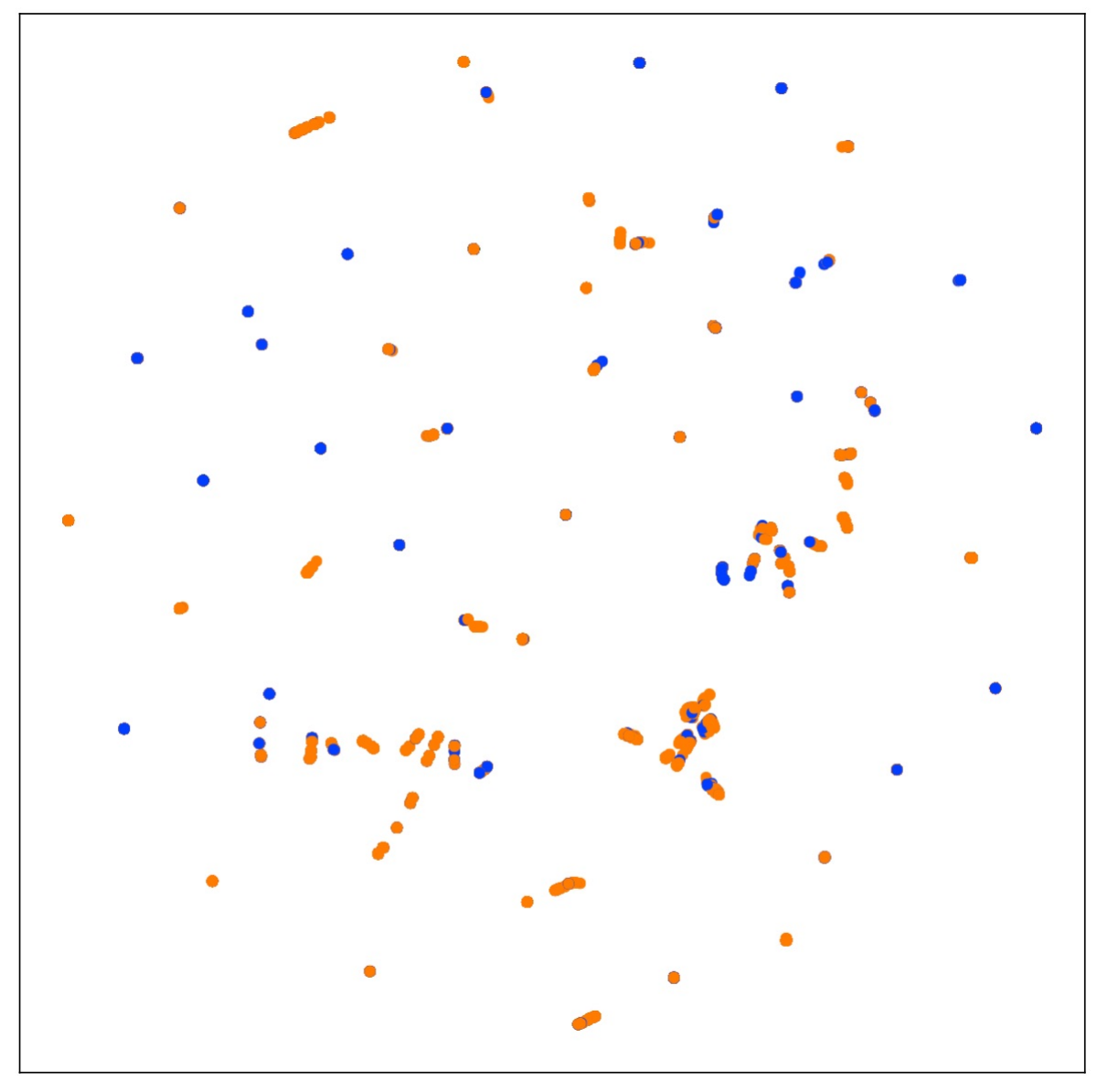}}
	\subfloat[{Loans: Learnt rep.}]
	{\centering \includegraphics[scale=0.20]{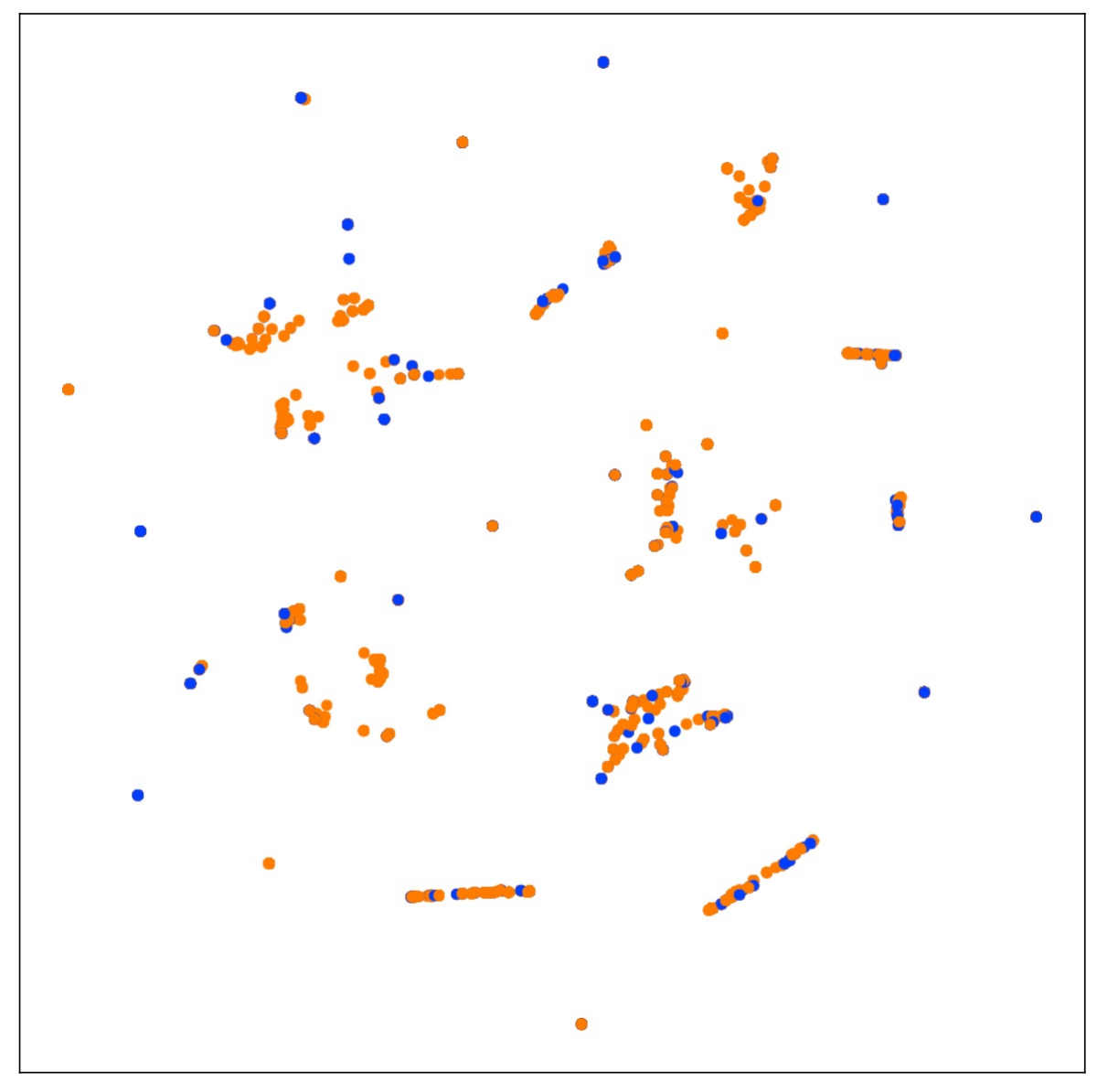}}
	\vspace{-2mm}
	\caption{2D tSNE of learnt representations by TAPPFL and of the raw input on a random client. Color means attribute value.} 
	\label{fig:TSNE_Adult}
	\vspace{-5mm}
\end{figure*}

\subsection{Experimental results}

\noindent {\bf Utility-privacy tradeoff.} According to Equation (\ref{eq:final-obj}), when $\lambda=0$ the first term of the objective function is disregarded, meaning that the protection of the private attribute is not considered. On the contrary, the second term is disappeared when $\lambda=1$, or in other words, we only consider protecting the private attribute and utility is not preserved. Our goal is to achieve a better trade-off by tuning $\lambda$ within $[0, 1]$, which allows preserving the FL utility and protecting the attribute privacy at the same time. 
Table~\ref{tab:allresults} shows the testing accuracy and average attribute inference accuracy of all devices in the considered datasets, where we set five different $\lambda$ values, i.e., 0, 0.25, 0.5, 0.75, and 1.0. 
We also show the gap between the attribute inference accuracy and the random guessing. 
The smaller the gap, the better the privacy protection. Ideally, when there is no gap, the learnt representation by our TAPPFL does not allow the
adversary (i.e., the server) to infer \emph{any} information related to the private attribute. 
Specifically, we have the following observations: 1) The testing accuracy is the largest when $\lambda=0$, hence the utility is maintained the most. However, the attribute inference accuracy is also the highest, indicating leaking the most attribute privacy.  
2) The attribute inference accuracy is the closest to random guessing when $\lambda=1$, meaning the attribute privacy is protected the most. However, the testing accuracy is also the smallest, indicating the utility is not well maintained. 
3) When $0<\lambda<1$, our TAPPFL achieves both reasonable testing accuracy and attribute inference accuracy---This indicates TAPPFL has a better utility-privacy tradeoff. 
Note that our TAPPFL does not know primary task's labels and learns the task-agnostic representations for device data during the entire training.

\vspace{+0.01in}
\noindent {\bf Mutual information vs. tradeoff parameter $\lambda$.} Further, we analyze our TAPPFL via plotting the two MI scores (i.e., the CE loss associated with {\bf Goal 1 (privacy protection)} and JSD loss associated with {\bf Goal 2 (utility preservation)}) vs. $\lambda$. Note that the CE 
and JSD loss are inversely proportional to the MI in the two goals.  
Figure~\ref{fig:plotMIs} shows the results on the three datasets. 
Each point indicates either a  CE loss or JSD loss at a selected $\lambda$. 
The tendency of these scores in function of $\lambda$ is presented by a trend line, which is computed using a least squares polynomial fit of first degree.
We observe that: 1) When the trade-off parameter $\lambda$ is low, the privacy protection is not carefully considered, which is translated into a high MI between the learnt representation and the private attribute, thus the  CE loss is relatively small. 
On the other hand, the utility is well-preserved, resulting also into a high MI between the input data and the learnt representation conditioned on the private attribute. So the JSD loss is relatively small. 
2)  Contrarily, 
for high values of $\lambda$, the privacy is largely protected in exchange for a large utility loss. 
Specifically, as $\lambda$ increases, the CE between the private attribute and the learnt representation increases, which is translated into a decrease of their MI, thus better protecting attribute privacy. Though not easily appreciated in the curves, the JSD loss tends to increase, 
thus reducing the utility.

\vspace{+0.01in}
\noindent {\bf Visualization of the learnt representations.}
In this experiment, we leverage the t-SNE embedding algorithm~\cite{van2008visualizing} to visualize the learnt representations by our trained feature extractor for the device data, and those without  our feature extractor. $\lambda$ is chosen in Table~\ref{tab:allresults} that achieves the best utility-privacy tradeoff. 
Figure~\ref{fig:TSNE_Adult} shows the 2D T-SNE visualization results, where each color corresponds to a private attribute value. We can observe that the 2D T-SNE embeddings of the raw input data form some clusters for the private attributes, meaning the private attributes can be easily inferred, e.g., the t-SNE representations via training a multi-class classifier. On the contrary, the 2D T-SNE embeddings of the learnt representations by our TAPPFL for different attribute values are completely mixed,  which thus makes it difficult for a malicious server to infer the private attributes from the learnt representations.

\begin{figure}[!t]
\vspace{-2mm}
\centering
\includegraphics[scale=0.3]{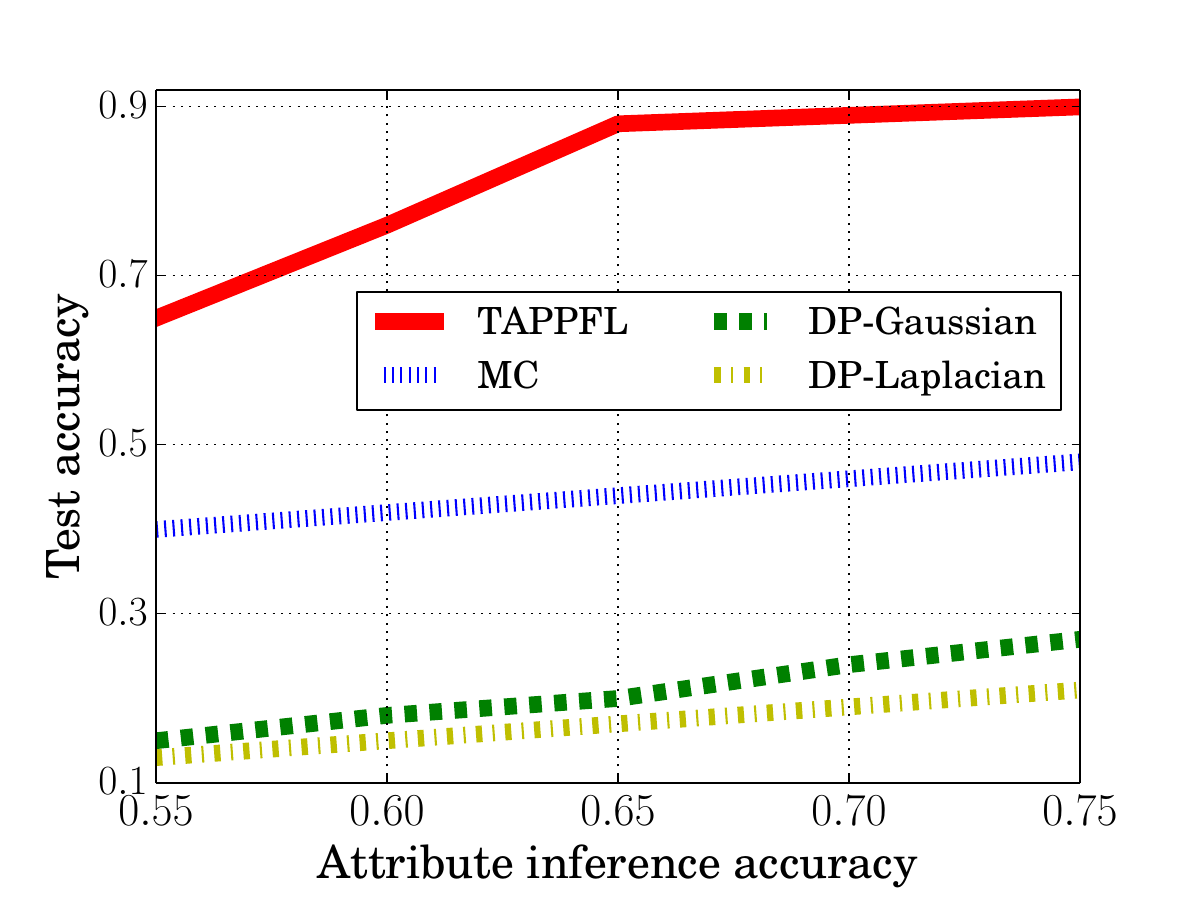}
\vspace{-3mm}
\caption{\small Compared defense results on CIFAR10.}
\vspace{-6mm}
\label{fig:defense_res}
\end{figure}

\vspace{+0.01in}
\noindent {\bf Comparison with task-agnostic defenses.}
In this experiment, we compare our TAPPFL with the two   \emph{task-agnostic} privacy protection methods, i.e., differential privacy (DP)~\cite{wei2020federated} and model compression (MC)~\cite{zhu2019deep} (See Table~\ref{tb:dpfl}). 
MC prunes the devices' feature extractor parameters whose magnitude are smaller than a threshold, and the devices only share parameters larger than the threshold to the server. DP ensures provable privacy guarantees \cite{WangSFSH22}. Specifically, DP randomly injects noise into the feature extractor's parameters and uploads the noisy parameters to the server. The server then performs the aggregation using the noisy parameters. Here, we consider applying the Gaussian noise and Laplacian noise to develop two DP baselines \cite{MohammadyXHZ0PD20}, i.e., DP-Gaussian and DP-Laplace (note that the DP protection in \cite{wei2020federated} is very weak due to very high $\epsilon$ such as 50 and 100). Thus, we tune the hyperparameter, i.e., noise variance in DP and pruning rate in MC, such that DP and MC have the same attribute inference accuracy as TAPPFL, and then compare their utility (i.e., testing accuracy). 
Figure~\ref{fig:defense_res} shows the comparison results on CIFAR10, where we set five attribute inference accuracy as 0.55, 0.60, 0.65, 0.70, and 0.75, respectively. 
We can observe that TAPPFL is significantly better than them. 
\emph{Note that we also evaluate DP by injecting noises to the final/input layer (as other existing methods did) and observe that DP-Gaussian and DP-Laplacian have very close (bad) performance as those injecting noises into the feature extractor's parameters. We do not show these results for simplicity.}

\section{Discussion}

{\bf Comparison with task-specific defenses.}
We also compare TAPPFL with 
state-of-the-art task-specific defenses, i.e.,  DPFE~\cite{osia2018deep} and adversarial training (AT)~\cite{li2019deepobfuscator} and show results on Loans under the same setting. 
\emph{Note that DPFE and  AT know the primary task.} 
We have the following observations: 1) Comparing with DPFE, with a same test accuracy as 0.84, TAPPFL's attribute inference accuracy is 0.63, while DPFE's is 0.77---showing a much worse protection than TAPPFL. One key reason is that DPFE estimates MI by assuming random variables to be Gaussian, which is inaccurate. Instead, TAPPFL does not have any assumption on the distributions of the random variables, meaning it can handle \emph{arbitrary} random variable distributions. 2) Comparing with AT, our TAPPFL's best-tradeoff is (test accuracy, inference accuracy) = (0.84, 0.63), while AT's is  (0.86, 0.63)---slightly better than ours. This is because both TAPPFL and AT involve adversarial training, but AT uses primary \emph{task labels}, while ours does not.

\vspace{+0.01in}
\noindent {\bf Defending against the state-of-the-art de-censoring representation attacks.} 
\citet{song2020overlearning} designed a de-censoring representation attack to better infer private attributes. Here, we test our TAPPFL against this attack on Loans. Specifically, by applying TAPPFL, when the test accuracy is 0.84, the inference accuracy \emph{without} \citet{song2020overlearning} is 0.63, while \emph{with} \citet{song2020overlearning} is 0.73, showing 
its better attack effectiveness. We then enhance privacy protection until the inference accuracy without \citet{song2020overlearning} is 0.57. In this case, its inference accuracy reduces to 0.60. The results show \citet{song2020overlearning}'s  performance could be largely reduced with more privacy protection imposed by TAPPFL. The fundamental reason is that TAPPFL has privacy guarantees against the (worst-case) inference attack.

\section{Conclusion}
\label{sec:conclusion}
We study privacy-preserving federated learning (FL) against the  attribute inference attack, i.e., an honest-but-curious server infers sensitive information in the device data from shared device models. 
To this end, we design a task-agnostic and provably privacy-preserving representation learning framework for FL (termed TAPPFL) from the information-theoretic perspective. 
TAPPFL is formulated via two mutual information goals:
one goal learns low-dimensional representations for device data that contain the least information
about the data's private attribute, and the other one includes
as much information as possible about the raw data, in order to maintain FL utility. 
TAPPFL also has upper bounded privacy leakage of the private attributes. 
Extensive results on various datasets from different applications show that 
TAPPFL can well protect the attributes (i.e., attribute inference accuracy is close to random guessing), and obtains a high utility. 
TAPPFL is also shown to largely outperform the state-of-the-art task-agnostic defenses.

\section*{\bf Acknowledgement} We thank the anonymous reviewers for their constructive feedback. 
Wang is partially supported by the Cisco Research Award, National Science Foundation under grant Nos. ECCS-2216926 and CNS-2241713. Hong is partially supported by the National Science Foundation under grant Nos. CNS-2302689 and CNS-2308730.  Any opinions, findings and conclusions or recommendations expressed in this material are those of the author(s) and do not necessarily reflect the views of the funding agencies.

{
\bibliography{refs,reference}

\begin{thebibliography}{59}
\providecommand{\natexlab}[1]{#1}

\bibitem[{Alemi et~al.(2017)Alemi, Fischer, Dillon, and Murphy}]{alemi2017deep}
Alemi, A.~A.; Fischer, I.; Dillon, J.~V.; and Murphy, K. 2017.
\newblock Deep Variational Information Bottleneck.
\newblock In \emph{ICLR}.

\bibitem[{{Alibaba Federated Learning}(2022)}]{AliFL}
{Alibaba Federated Learning}. 2022.
\newblock \url{https://federatedscope.io/}.

\bibitem[{Aono et~al.(2017)Aono, Hayashi, Wang, and Moriai}]{aono2017privacy}
Aono, Y.; Hayashi, T.; Wang, L.; and Moriai, S. 2017.
\newblock Privacy-preserving deep learning: Revisited and enhanced.
\newblock In \emph{ATIS}.

\bibitem[{Belghazi et~al.(2018)Belghazi, Baratin, Rajeshwar, Ozair, Bengio, Courville, and Hjelm}]{belghazi2018mutual}
Belghazi, M.~I.; Baratin, A.; Rajeshwar, S.; Ozair, S.; Bengio, Y.; Courville, A.; and Hjelm, D. 2018.
\newblock Mutual information neural estimation.
\newblock In \emph{ICML}.

\bibitem[{Bonawitz et~al.(2019)Bonawitz, Eichner, Grieskamp, and et~al.}]{bonawitz2019towards}
Bonawitz, K.; Eichner, H.; Grieskamp, W.; and et~al. 2019.
\newblock Towards federated learning at scale: System design.
\newblock \emph{MLSys}.

\bibitem[{Bonawitz et~al.(2017)Bonawitz, Ivanov, Kreuter, Marcedone, McMahan, Patel, Ramage, Segal, and Seth}]{bonawitz2017practical}
Bonawitz, K.; Ivanov, V.; Kreuter, B.; Marcedone, A.; McMahan, H.~B.; Patel, S.; Ramage, D.; Segal, A.; and Seth, K. 2017.
\newblock Practical secure aggregation for privacy-preserving machine learning.
\newblock In \emph{CCS}.

\bibitem[{Calabro(2009)}]{calabro2009exponential}
Calabro, C. 2009.
\newblock \emph{The exponential complexity of satisfiability problems}.
\newblock University of California, San Diego.

\bibitem[{Chen et~al.(2016)Chen, Duan, Houthooft, Schulman, Sutskever, and Abbeel}]{chen2016infogan}
Chen, X.; Duan, Y.; Houthooft, R.; Schulman, J.; Sutskever, I.; and Abbeel, P. 2016.
\newblock Infogan: Interpretable representation learning by information maximizing generative adversarial nets.
\newblock In \emph{NIPS}.

\bibitem[{Cheng et~al.(2020)Cheng, Hao, Dai, Liu, Gan, and Carin}]{cheng2020club}
Cheng, P.; Hao, W.; Dai, S.; Liu, J.; Gan, Z.; and Carin, L. 2020.
\newblock CLUB: A Contrastive Log-ratio Upper Bound of Mutual Information.
\newblock In \emph{ICML}.

\bibitem[{Dang et~al.(2021)Dang, Thakkar, Ramaswamy, Mathews, Chin, and Beaufays}]{dang2021revealing}
Dang, T.; Thakkar, O.; Ramaswamy, S.; Mathews, R.; Chin, P.; and Beaufays, F. 2021.
\newblock Revealing and protecting labels in distributed training.
\newblock In \emph{NeurIPS}.

\bibitem[{Danner and Jelasity(2015)}]{danner2015fully}
Danner, G.; and Jelasity, M. 2015.
\newblock Fully distributed privacy preserving mini-batch gradient descent learning.
\newblock In \emph{IFIP DAIS}.

\bibitem[{Dua and Graff(2017)}]{Dua:2019}
Dua, D.; and Graff, C. 2017.
\newblock {UCI} Machine Learning Repository.

\bibitem[{Ganju et~al.(2018)Ganju, Wang, Yang, Gunter, and Borisov}]{ganju2018property}
Ganju, K.; Wang, Q.; Yang, W.; Gunter, C.~A.; and Borisov, N. 2018.
\newblock Property inference attacks on fully connected neural networks using permutation invariant representations.
\newblock In \emph{CCS}.

\bibitem[{Geyer, Klein, and Nabi(2017)}]{geyer2017differentially}
Geyer, R.~C.; Klein, T.; and Nabi, M. 2017.
\newblock Differentially private federated learning: A client level perspective.
\newblock \emph{arXiv}.

\bibitem[{Gibbs and Su(2002)}]{gibbs2002choosing}
Gibbs, A.~L.; and Su, F.~E. 2002.
\newblock On choosing and bounding probability metrics.
\newblock \emph{International statistical review}, 70(3): 419--435.

\bibitem[{Goodfellow et~al.(2014)Goodfellow, Pouget-Abadie, Mirza, Xu, Warde-Farley, Ozair, Courville, and Bengio}]{goodfellow2014generative}
Goodfellow, I.; Pouget-Abadie, J.; Mirza, M.; Xu, B.; Warde-Farley, D.; Ozair, S.; Courville, A.; and Bengio, Y. 2014.
\newblock Generative adversarial nets.
\newblock In \emph{NIPS}.

\bibitem[{{Google Federated Learning}(2022)}]{GoogleFL}
{Google Federated Learning}. 2022.
\newblock \url{https://federated.withgoogle.com/}.

\bibitem[{Hamm, Cao, and Belkin(2016)}]{hamm2016learning}
Hamm, J.; Cao, Y.; and Belkin, M. 2016.
\newblock Learning privately from multiparty data.
\newblock In \emph{ICML}.

\bibitem[{Hardt, Price, and Srebro(2016)}]{10.5555/3157382.3157469}
Hardt, M.; Price, E.; and Srebro, N. 2016.
\newblock Equality of Opportunity in Supervised Learning.

\bibitem[{Hjelm et~al.(2019)Hjelm, Fedorov, Lavoie-Marchildon, Grewal, Bachman, Trischler, and Bengio}]{hjelm2019learning}
Hjelm, R.~D.; Fedorov, A.; Lavoie-Marchildon, S.; Grewal, K.; Bachman, P.; Trischler, A.; and Bengio, Y. 2019.
\newblock Learning deep representations by mutual information estimation and maximization.
\newblock In \emph{ICLR}.

\bibitem[{{IBM Federated Learning}(2022)}]{IBMFL}
{IBM Federated Learning}. 2022.
\newblock \url{https://www.ibm.com/docs/en/cloud-paks/cp-data/4.0?topic=models-federated-learning}.

\bibitem[{Jia et~al.(2017)Jia, Wang, Zhang, and Gong}]{jia2017attriinfer}
Jia, J.; Wang, B.; Zhang, L.; and Gong, N.~Z. 2017.
\newblock AttriInfer: Inferring user attributes in online social networks using markov random fields.
\newblock In \emph{WWW}.

\bibitem[{Kaissis et~al.(2020)Kaissis, Makowski, R{\"u}ckert, and Braren}]{kaissis2020secure}
Kaissis, G.~A.; Makowski, M.~R.; R{\"u}ckert, D.; and Braren, R.~F. 2020.
\newblock Secure, privacy-preserving and federated machine learning in medical imaging.
\newblock \emph{Nature Machine Intelligence}.

\bibitem[{Keahey et~al.(2020)Keahey, Anderson, Zhen, Riteau, Ruth, Stanzione, Cevik, Colleran, Gunawi, Hammock, Mambretti, Barnes, Halbach, Rocha, and Stubbs}]{keahey2020lessons}
Keahey, K.; Anderson, J.; Zhen, Z.; Riteau, P.; Ruth, P.; Stanzione, D.; Cevik, M.; Colleran, J.; Gunawi, H.~S.; Hammock, C.; Mambretti, J.; Barnes, A.; Halbach, F.; Rocha, A.; and Stubbs, J. 2020.
\newblock Lessons Learned from the Chameleon Testbed.
\newblock In \emph{USENIX ATC}.

\bibitem[{Kim et~al.(2019)Kim, Kang, Pulli, and Choi}]{kim2019training}
Kim, T.-h.; Kang, D.; Pulli, K.; and Choi, J. 2019.
\newblock Training with the invisibles: Obfuscating images to share safely for learning visual recognition models.
\newblock \emph{arXiv}.

\bibitem[{Kingma, Welling et~al.(2019)}]{kingma2019introduction}
Kingma, D.~P.; Welling, M.; et~al. 2019.
\newblock An introduction to variational autoencoders.
\newblock \emph{Foundations and Trends{\textregistered} in Machine Learning}, 12(4): 307--392.

\bibitem[{Krizhevsky(2009)}]{Krizhevsky09learningmultiple}
Krizhevsky, A. 2009.
\newblock Learning multiple layers of features from tiny images.
\newblock Technical report.

\bibitem[{Li et~al.(2020{\natexlab{a}})Li, Guo, Yang, and Chen}]{li2019deepobfuscator}
Li, A.; Guo, J.; Yang, H.; and Chen, Y. 2020{\natexlab{a}}.
\newblock Deepobfuscator: Adversarial training framework for privacy-preserving image classification.
\newblock \emph{arXiv preprint arXiv:1909.04126}.

\bibitem[{Li et~al.(2020{\natexlab{b}})Li, Sahu, Talwalkar, and Smith}]{li2020federated}
Li, T.; Sahu, A.~K.; Talwalkar, A.; and Smith, V. 2020{\natexlab{b}}.
\newblock Federated learning: Challenges, methods, and future directions.
\newblock \emph{IEEE Signal Processing Magazine}.

\bibitem[{Liao et~al.(2021)Liao, Zhao, Xu, Jaakkola, Gordon, Jegelka, and Salakhutdinov}]{liao2021information}
Liao, P.; Zhao, H.; Xu, K.; Jaakkola, T.; Gordon, G.~J.; Jegelka, S.; and Salakhutdinov, R. 2021.
\newblock Information obfuscation of graph neural networks.
\newblock In \emph{ICML}.

\bibitem[{Liu et~al.(2019)Liu, Du, Shrivastava, and Zhong}]{liu2019privacy}
Liu, S.; Du, J.; Shrivastava, A.; and Zhong, L. 2019.
\newblock Privacy Adversarial Network: Representation Learning for Mobile Data Privacy.
\newblock \emph{Proceedings of the ACM on Interactive, Mobile, Wearable and Ubiquitous Technologies}, 3(4): 1--18.

\bibitem[{Madras et~al.(2018)Madras, Creager, Pitassi, and Zemel}]{madras2018learning}
Madras, D.; Creager, E.; Pitassi, T.; and Zemel, R. 2018.
\newblock Learning adversarially fair and transferable representations.
\newblock In \emph{ICML}.

\bibitem[{McMahan et~al.(2017{\natexlab{a}})McMahan, Moore, Ramage, Hampson, and y~Arcas}]{mcmahan2017communication}
McMahan, B.; Moore, E.; Ramage, D.; Hampson, S.; and y~Arcas, B.~A. 2017{\natexlab{a}}.
\newblock Communication-Efficient Learning of Deep Networks from Decentralized Data.
\newblock In \emph{AISTATS}.

\bibitem[{McMahan et~al.(2017{\natexlab{b}})McMahan, Moore, Ramage, Hampson, and y~Arcas}]{McMahan17}
McMahan, H.~B.; Moore, E.; Ramage, D.; Hampson, S.; and y~Arcas, B.~A. 2017{\natexlab{b}}.
\newblock Communication-Efficient Learning of Deep Networks from Decentralized Data.
\newblock In \emph{AISTATS}.

\bibitem[{McMahan et~al.(2018)McMahan, Ramage, Talwar, and Zhang}]{mcmahan2018learning}
McMahan, H.~B.; Ramage, D.; Talwar, K.; and Zhang, L. 2018.
\newblock Learning Differentially Private Recurrent Language Models.
\newblock In \emph{ICLR}.

\bibitem[{Melis et~al.(2019)Melis, Song, De~Cristofaro, and Shmatikov}]{melis2019exploiting}
Melis, L.; Song, C.; De~Cristofaro, E.; and Shmatikov, V. 2019.
\newblock Exploiting unintended feature leakage in collaborative learning.
\newblock In \emph{IEEE SP}.

\bibitem[{{Microsoft Federated Learning}(2022)}]{MSFL}
{Microsoft Federated Learning}. 2022.
\newblock \url{https://www.microsoft.com/en-us/research/blog/flute-a-scalable-federated-learning-simulation-platform/}.

\bibitem[{Mohammady et~al.(2020)Mohammady, Xie, Hong, Zhang, Wang, Pourzandi, and Debbabi}]{MohammadyXHZ0PD20}
Mohammady, M.; Xie, S.; Hong, Y.; Zhang, M.; Wang, L.; Pourzandi, M.; and Debbabi, M. 2020.
\newblock {R2DP:} {A} Universal and Automated Approach to Optimizing the Randomization Mechanisms of Differential Privacy for Utility Metrics with No Known Optimal Distributions.
\newblock In \emph{CCS}, 677--696. {ACM}.

\bibitem[{Mohassel and Zhang(2017)}]{mohassel2017secureml}
Mohassel, P.; and Zhang, Y. 2017.
\newblock Secureml: A system for scalable privacy-preserving machine learning.
\newblock In \emph{IEEE SP}.

\bibitem[{Moyer et~al.(2018)Moyer, Gao, Brekelmans, Galstyan, and Ver~Steeg}]{moyer2018invariant}
Moyer, D.; Gao, S.; Brekelmans, R.; Galstyan, A.; and Ver~Steeg, G. 2018.
\newblock Invariant representations without adversarial training.
\newblock In \emph{NeurIPS}.

\bibitem[{Oh, Fritz, and Schiele(2017)}]{oh2017adversarial}
Oh, S.~J.; Fritz, M.; and Schiele, B. 2017.
\newblock Adversarial image perturbation for privacy protection a game theory perspective.
\newblock In \emph{ICCV}.

\bibitem[{Oord, Li, and Vinyals(2018)}]{oord2018representation}
Oord, A. v.~d.; Li, Y.; and Vinyals, O. 2018.
\newblock Representation learning with contrastive predictive coding.
\newblock \emph{arXiv}.

\bibitem[{Osia et~al.(2018)Osia, Taheri, Shamsabadi, Katevas, Haddadi, and Rabiee}]{osia2018deep}
Osia, S.~A.; Taheri, A.; Shamsabadi, A.~S.; Katevas, K.; Haddadi, H.; and Rabiee, H.~R. 2018.
\newblock Deep private-feature extraction.
\newblock \emph{IEEE TKDE}.

\bibitem[{Pathak, Rane, and Raj(2010)}]{pathak2010multiparty}
Pathak, M.; Rane, S.; and Raj, B. 2010.
\newblock Multiparty differential privacy via aggregation of locally trained classifiers.
\newblock In \emph{NIPS}.

\bibitem[{Peng et~al.(2018)Peng, Kanazawa, Toyer, Abbeel, and Levine}]{peng2018variational}
Peng, X.~B.; Kanazawa, A.; Toyer, S.; Abbeel, P.; and Levine, S. 2018.
\newblock Variational discriminator bottleneck: Improving imitation learning, inverse rl, and gans by constraining information flow.
\newblock \emph{arXiv preprint arXiv:1810.00821}.

\bibitem[{Pittaluga, Koppal, and Chakrabarti(2019)}]{pittaluga2019learning}
Pittaluga, F.; Koppal, S.; and Chakrabarti, A. 2019.
\newblock Learning privacy preserving encodings through adversarial training.
\newblock In \emph{WACV}.

\bibitem[{Poole et~al.(2019)Poole, Ozair, Oord, Alemi, and Tucker}]{poole2019variational}
Poole, B.; Ozair, S.; Oord, A. v.~d.; Alemi, A.~A.; and Tucker, G. 2019.
\newblock On variational bounds of mutual information.
\newblock In \emph{ICML}.

\bibitem[{Rieke et~al.(2020)Rieke, Hancox, Li, Milletari, Roth, Albarqouni, Bakas, Galtier, Landman, Maier-Hein et~al.}]{rieke2020future}
Rieke, N.; Hancox, J.; Li, W.; Milletari, F.; Roth, H.~R.; Albarqouni, S.; Bakas, S.; Galtier, M.~N.; Landman, B.~A.; Maier-Hein, K.; et~al. 2020.
\newblock The future of digital health with federated learning.
\newblock \emph{NPJ digital medicine}, 3(1): 119.

\bibitem[{Shokri and Shmatikov(2015)}]{shokri2015privacy}
Shokri, R.; and Shmatikov, V. 2015.
\newblock Privacy-preserving deep learning.
\newblock In \emph{CCS}.

\bibitem[{Song and Shmatikov(2020)}]{song2020overlearning}
Song, C.; and Shmatikov, V. 2020.
\newblock Overlearning Reveals Sensitive Attributes.
\newblock In \emph{ICLR}.

\bibitem[{Song et~al.(2019)Song, Kalluri, Grover, Zhao, and Ermon}]{song2019learning}
Song, J.; Kalluri, P.; Grover, A.; Zhao, S.; and Ermon, S. 2019.
\newblock Learning controllable fair representations.
\newblock In \emph{AISTATS}.

\bibitem[{Van~der Maaten and Hinton(2008)}]{van2008visualizing}
Van~der Maaten, L.; and Hinton, G. 2008.
\newblock Visualizing data using t-SNE.
\newblock \emph{JMLR}.

\bibitem[{Wainakh et~al.(2022)Wainakh, Ventola, M{\"u}{\ss}ig, Keim, Cordero, Zimmer, Grube, Kersting, and M{\"u}hlh{\"a}user}]{wainakh2022user}
Wainakh, A.; Ventola, F.; M{\"u}{\ss}ig, T.; Keim, J.; Cordero, C.~G.; Zimmer, E.; Grube, T.; Kersting, K.; and M{\"u}hlh{\"a}user, M. 2022.
\newblock User-Level Label Leakage from Gradients in Federated Learning.
\newblock In \emph{PTES}.

\bibitem[{Wang et~al.(2021)Wang, Guo, Li, Chen, and Li}]{wang2021privacy}
Wang, B.; Guo, J.; Li, A.; Chen, Y.; and Li, H. 2021.
\newblock Privacy-Preserving Representation Learning on Graphs: A Mutual Information Perspective.
\newblock In \emph{KDD}.

\bibitem[{Wang et~al.(2022)Wang, Sharma, Feng, Shu, and Hong}]{WangSFSH22}
Wang, H.; Sharma, J.; Feng, S.; Shu, K.; and Hong, Y. 2022.
\newblock A Model-Agnostic Approach to Differentially Private Topic Mining.
\newblock In \emph{KDD}, 1835--1845. {ACM}.

\bibitem[{Wei et~al.(2020)Wei, Li, Ding, Ma, Yang, Farokhi, Jin, Quek, and Poor}]{wei2020federated}
Wei, K.; Li, J.; Ding, M.; Ma, C.; Yang, H.~H.; Farokhi, F.; Jin, S.; Quek, T.~Q.; and Poor, H.~V. 2020.
\newblock Federated learning with differential privacy: Algorithms and performance analysis.
\newblock \emph{IEEE TIFS}.

\bibitem[{Wu et~al.(2018)Wu, Wang, Wang, and Jin}]{wu2018towards}
Wu, Z.; Wang, Z.; Wang, Z.; and Jin, H. 2018.
\newblock Towards privacy-preserving visual recognition via adversarial training: A pilot study.
\newblock In \emph{ECCV}.

\bibitem[{Zhao et~al.(2020)Zhao, Chi, Tian, and Gordon}]{zhao2020trade}
Zhao, H.; Chi, J.; Tian, Y.; and Gordon, G.~J. 2020.
\newblock Trade-offs and guarantees of adversarial representation learning for information obfuscation.
\newblock In \emph{NeurIPS}.

\bibitem[{Zhu, Liu, and Han(2019)}]{zhu2019deep}
Zhu, L.; Liu, Z.; and Han, S. 2019.
\newblock Deep leakage from gradients.
\newblock In \emph{NeurIPS}.

\end{thebibliography}
}

\newpage
\clearpage 
\appendix

\begin{algorithm}[t]
\small
\caption{Task-agnostic privacy-preserving rep. learning for FL against attribute inference attacks ({\bf TAPPFL)}}
\begin{algorithmic}[1]
    \REQUIRE $\rho$: fraction of participating devices per round; $\mathcal{C}=\{C_i\}_{i=1}^M$: $M$ total devices; 
    $B$: batch size; $E$: \#local epochs; $T$: \#global rounds; 
    $lr_1, lr_2, lr_3$: learning rates; 
    $\lambda \in [0,1]$: utility-privacy tradeoff parameter. 
    \ENSURE $\{\Theta_i^T\}_{i=1}^M$, $\{\Psi_i^T\}_{i=1}^M$, $\{\Omega_i^T\}_{i=1}^M$

    \STATE \textbf{Initialization:} $\{\Theta_i^0, \Psi_i^0, \Omega_i^0\ \}_{i=1}^M$.
    E.g.,  $\{\Theta_i^0\}_{i=1}^M$ are initialized via  pretraining each feature extractor NN. 
    
    \FOR {global round $t=0,1,2, \cdots, T-1$}
        \FOR{each device $C_i \in  \mathcal{C}$}
            \STATE $\Theta_{i}^{t+1} \gets$ $\textbf{DeviceUpdate}(i, \Theta_{i}^t)$
        \ENDFOR
        
        \STATE $\Theta^{t+1} \gets$ $\textbf{ServerUpdate}(\{\Theta_{i}^{t+1}\}, \rho)$
        \STATE Set $\{\Theta_{i}^{t+1}\} \gets \Theta^{t+1} $
	\ENDFOR
	\\ 
	
    \STATE $\textbf{DeviceUpdate}(i, \Theta^t):$

     
     \STATE $\Theta_i^t \gets \Theta^t$

     \STATE $CE\_loss= CE(u^i,g_{\Psi^i}(f_{\Theta_i^t}(\mathbf{x}^i)))$
    \STATE $JSD\_loss= -I^{(JSD)} (\mathbf{x}^i,f_{\Theta_i^t}(\mathbf{x}^i),u^i)$
    \FOR {local epoch $e=1,2, \cdots, E$}
         \STATE $\mathcal{B} \gets$ Split device $C_i$'s data into mini-batches of size $B$
         
        \FOR{each min-batch $b \in \mathcal{B}$}
        \STATE $\Psi_i^{t+1} \gets \Psi_i^t - lr_1  \cdot {\partial{CE\_loss}}/{\partial{\Psi_i^t}}$
        \STATE $\Omega_i^{t+1} \gets \Omega_i^t { - } lr_2  \cdot {\partial{JSD\_loss}}/{\partial{\Omega_i^t}}$
        \STATE $\Theta_i^{t+1} \gets \Theta_i^{t} { - } lr_3 \cdot { \partial{\big(\lambda CE\_loss +(1-\lambda)  JSD\_loss\big)}}/{\partial{\Theta_i^t}}$
        \ENDFOR
    \ENDFOR

\STATE $\textbf{ServerUpdate}(\{\Theta_{i}^{t+1}\}, \rho):$
\STATE $\mathcal{C}_K \gets$ randomly select $K=\rho \cdot M$ devices
\STATE $\Theta^{t+1} \gets \frac{1}{K}\sum_{k \in \mathcal{C}_K}  \Theta^{t+1}_{k}$
\end{algorithmic} 
\label{alg:tappfl}
\end{algorithm}

\section{Proofs}

\subsection{Proof of Theorem \ref{thm:uptradeoff}}
\label{supp:uptradeoff}
We first introduce the following definitions and lemmas that will be used to prove Theorem~\ref{thm:uptradeoff}.
\begin{definition}[Total variance (TV) distance]
\label{def:TV}
Let $\mathcal{D}_1$ and $\mathcal{D}_2$ be two distributions over the same sample space $\Gamma$,  the TV distance between $\mathcal{D}_1$ and $\mathcal{D}_2$ is defined as: $d_{TV}(\mathcal{D}_1, \mathcal{D}_2) =  \max_{E \subseteq \Gamma} |\mathcal{D}_1 (E) - \mathcal{D}_2(E)|$.   
\end{definition} 

\begin{definition}[1-Wasserstein distance]
\label{def:wassdis}
Let $\mathcal{D}_1$ and $\mathcal{D}_2$ be two distributions over the same sample space $\Gamma$, the 1-Wasserstein distance between $\mathcal{D}_1$ and $\mathcal{D}_2$ is defined as $W_1(\mathcal{D}_1, \mathcal{D}_2) = \max_{\|f\|_L \leq 1} |\int_{\Gamma} f d\mathcal{D}_1 - \int_{\Gamma} f d\mathcal{D}_2 |$, where $\|\cdot\|_L$  is the Lipschitz norm of a real-valued function. 
\end{definition}

\begin{definition}[Pushforward distribution]
\label{def:pushforDist}
Let $\mathcal{D}$ be a distribution over a sample space and $g$ be a function of the same space. Then we call $g(\mathcal{D})$ the pushforward distribution of $\mathcal{D}$.
\end{definition}

\begin{lemma}[Contraction of the 1-Wasserstein distance]
\label{lem:contWass}
Let $g$ be a function defined on a space and $L$ be constant such that $\|g\|_L \leq C_L$. 
For any distributions $\mathcal{D}_1$ and $\mathcal{D}_2$ over this space, 
$ W_1(g(\mathcal{D}_1), g(\mathcal{D}_2))  \leq C_L \cdot W_1(\mathcal{D}_1, \mathcal{D}_2)$.
\end{lemma}

\begin{lemma}[1-Wasserstein distance on Bernoulli random variables]
\label{lem:berprob}
Let $y_1$ and $y_2$ be two Bernoulli random variables with distributions $\mathcal{D}_1$ and $\mathcal{D}_2$, respectively. Then, 
$W_1(\mathcal{D}_1, \mathcal{D}_2)  = |\textrm{Pr}(y_1=1) - \textrm{Pr}(y_2=1)|$.
\end{lemma}

\begin{lemma}
[Relationship between the 1-Wasserstein distance and the TV distance~\cite{gibbs2002choosing}]
\label{lem:tvWass}
Let $g$ be a function defined on a norm-bounded space $\mathcal{Z}$, where $\max_{{\bf r} \in \mathcal{Z}} \|{\bf r} \| \leq R$, and $\mathcal{D}_1$ and $\mathcal{D}_1$ are two distributions over the space $\mathcal{Z}$.  Then $W_1(g(\mathcal{D}_1), g(\mathcal{D}_2)) \leq 2R \cdot d_{TV}(g(\mathcal{D}_1), g(\mathcal{D}_2))$.  
\end{lemma}

We now prove Theorem~\ref{thm:uptradeoff}, which is restated as below: 
\uptradeoff*

\begin{proof}
We denote $\mathcal{D}^i_{u^i=a}$ as the conditional distribution of $\mathcal{D}^i$ given $u^i=a$, and $cf_i$ as the (binary) composition function of  $c \circ f_{\Theta_i}$. 
As $c$ is binary task classifier on the learnt representations, it follows that the pushforward 
$cf_i(\mathcal{D}^i_{u^i=a})$ induces two distributions over $\{0,1\}$ with $a=\{0,1\}$. 

By leveraging the triangle inequalities of the 1-Wasserstein distance, we have
\begin{small}
\begin{align}
& W_1 (\mathcal{D}^i_{{\bf y}^i|u^i=0}, \mathcal{D}^i_{{\bf y}^i|u^i=1}) \nonumber \\ 
& \leq W_1 (\mathcal{D}^i_{{\bf y}^i|u^i=0}, cf_i(\mathcal{D}^i_{u^i=0})) + 
W_1 (cf_i(\mathcal{D}^i_{u^i=0}), cf_i(\mathcal{D}^i_{u^i=1})) \notag \\
& \quad + W_1 (cf_i(\mathcal{D}^i_{u^i=1}), \mathcal{D}^i_{{\bf y}^i|u^i=1})
\label{eqn:keytri}
\end{align}
\end{small}

Using Lemma~\ref{lem:berprob} on Bernoulli random variables ${\bf y}^i|u^i=a$, we have 
\begin{small}
\begin{align}
\label{eqn:berprob}
& W_1 (\mathcal{D}^i_{{\bf y}^i|u^i=0}, \mathcal{D}^i_{{\bf y}^i|u^i=1}) \notag \\
& = |\textrm{Pr}_{\mathcal{D}^i}({\bf y}^i=1|u^i=0) - \textrm{Pr}_{\mathcal{D}^i}({\bf y}^i=1|u^i=1) | \notag \\
& = \Delta_{{\bf y}^i|u^i}.
\end{align}
\end{small}

Using Lemma~\ref{lem:contWass} on the contraction of the 1-Wasserstein distance and that $\|c\|_L \leq C_L$, we have 
\begin{small}
\begin{align}
\label{eqn:contWass}
W_1 (cf_i(\mathcal{D}^i_{u^i=0}), cf_i(\mathcal{D}^i_{u^i=1})) \leq C_L \cdot W_1(f_i(\mathcal{D}^i_{u^i=0}), f_i(\mathcal{D}^i_{u^i=1})).
\end{align}
\end{small}

Using Lemma~\ref{lem:tvWass} with $\max_{i,{\bf r}^i} \|{\bf r}^i\| \leq R$, we have 
\begin{small}
\begin{align}
\label{eqn:tvWass}
W_1(f_i(\mathcal{D}^i_{u^i=0}), f_i(\mathcal{D}^i_{u^i=1})) \leq 2R \cdot d_{TV}(f_i(\mathcal{D}^i_{u^i=0}), f_i(\mathcal{D}^i_{u^i=1})). 
\end{align}
\end{small}%

We further show $d_{TV}(f_i(\mathcal{D}^i_{u^i=0}), f_i(\mathcal{D}^i_{u^i=1})) = \textrm{Adv}_{\mathcal{D}^i}(\mathcal{A})$, as proven in~\cite{liao2021information}. Specifically, 
\begin{small}
\begin{align}
& d_{TV}(f_i(\mathcal{D}^i_{u^i=0}), f_i(\mathcal{D}^i_{u^i=1})) \notag \\
& = \max_{E} |\textrm{Pr}_{f_i(\mathcal{D}^i_{u^i=0})}(E) - \textrm{Pr}_{f_i(\mathcal{D}^i_{u^i=1})}(E)| \nonumber \\
& = \max_{A_E \in \mathcal{A}} | \textrm{Pr}_{{\bf r}^i \sim f_i(\mathcal{D}^i_{u^i=0})}(A_E({\bf r}^i)=1) \notag \\ 
& \qquad - \textrm{Pr}_{{\bf r}^i \sim f_i(\mathcal{D}^i_{u^i=1})}(A_E({\bf r}^i)=1) | \nonumber \\
& = \max_{A_E \in \mathcal{A}} | \textrm{Pr}(A_E({\bf r}^i)=1 | u^i=0) - \textrm{Pr}(A_E({\bf r}^i)=1 | u^i=1) | \nonumber \\
& = \textrm{Adv}_{\mathcal{D}^i}(\mathcal{A}), \label{eqn:tvadv}
\end{align}
\end{small}%
where the first equation uses the definition of TV distance, and  $A_E(\cdot)$ is the characteristic function of the event $E$ in the second equation.

With Equations~\ref{eqn:contWass}-\ref{eqn:tvadv}, we have 
$$W_1 (cf_i(\mathcal{D}^i_{u^i=0}), cf_i(\mathcal{D}^i_{u^i=1})) \leq 2R \cdot  C_L \cdot \textrm{Adv}_{\mathcal{D}^i}(\mathcal{A}).$$ 
Furthermore, using Lemma~\ref{lem:berprob} on Bernoulli random variables ${\bf y}^i$ and $cf_i({\bf x}^i)$, we have 
\begin{align}
& W_1 (\mathcal{D}^i_{{\bf y}^i|u^i=a}, cf_i(\mathcal{D}^i_{u^i=a})) \notag \\
& = |\textrm{Pr}_{\mathcal{D}^i}({\bf y}^i=1 | u^i=a) - \textrm{Pr}_{\mathcal{D}^i}({cf_i({\bf x}^i})=1 | u^i=a)) | \nonumber \\
& = |\mathbb{E}_{\mathcal{D}^i}[{\bf y}^i|u^i=a] - \mathbb{E}_{\mathcal{D}^i}[cf_i({\bf x}^i)| u^i=a]| \nonumber \\
& \leq \mathbb{E}_{\mathcal{D}^i}[|{\bf y}^i-cf_i({\bf x}^i)| |u^i=a] \nonumber \\
& = \textrm{Pr}_{\mathcal{D}^i}({\bf y}^i \neq cf_i({\bf x}^i) | u^i=a) \nonumber \\
& \leq CE_{u^i=a}({\bf y}^i, cf_i({\bf x}^i)), \label{eqn:CEloss}
\end{align}
where we use the fact that cross-entropy loss is an upper bound of the binary loss in the last inequality. 

Finally, by combining Equation~\ref{eqn:contWass}-Equation~\ref{eqn:CEloss}, we have:  
\begin{align}
    & \Delta_{{\bf y}^i|u^i} \leq CE_{u^i=0}({\bf y}^i, cf_i({\bf x}^i)) + 2R \cdot C_L \cdot \textrm{Adv}_{\mathcal{D}^i}(\mathcal{A}) \notag \\
    & \qquad + CE_{u^i=1}({\bf y}^i, cf_i({\bf x}^i))
\end{align}

Hence, 
$
    \textrm{err}_i = 
    CE_{u^i=0} ({\bf y}^i, c({\bf r}^i)) + CE_{u^i=1} ({\bf y}^i, c({\bf r}^i)) 
    \geq \Delta_{{\bf y}^i|u^i} - 2R \cdot C_L \cdot \textrm{Adv}_{\mathcal{D}^i}(\mathcal{A}),
$ completing the proof. 

\end{proof}

\subsection{Proof of Theorem \ref{thm:provprivacy}}
\label{supp:provprivacy}

We note that \cite{zhao2020trade} also provide similar theoretical result in Theorem 3.1 against attribute inference attacks. However, there are two key differences between theirs and our Theorem~\ref{thm:provprivacy}: 
(i) Theorem 3.1  requires an assumption $I(\hat{A}; A|Z)=0$, while our Theorem 5 does not need extra assumption; (ii) The proof for Theorem 3.1 decomposes an joint entropy, while our proof decomposes a conditional entropy $H(s^i, u^i | \mathcal{A}({\bf r}^i))$. We note that the main idea is by introducing an indicator and decomposing an entropy in two different ways---which is a common strategy.

The following lemma about the inverse binary entropy will be useful in the proof of Theorem~\ref{thm:provprivacy}:
\begin{lemma}[\cite{calabro2009exponential} Theorem 2.2]
\label{lem:inveren}
Let $H_2^{-1}(p)$ be the inverse binary entropy function for $p \in [0,1]$, then $H_2^{-1}(p) \geq \frac{p}{2 \log_2(\frac{6}{p})}$.
\end{lemma}

\begin{lemma}[Data processing inequality] 
\label{lem:datainq}
Given random variables $X$, $Y$, and $Z$ that form a Markov chain in the
order $X \rightarrow Y \rightarrow Z$, then the mutual information between $X$ and $Y$ is greater than or equal to the mutual information between $X$ and $Z$. That is $I(X;Y) \geq I(X;Z)$.
\end{lemma}

With the above lemma, we are ready to prove Theorem~\ref{thm:provprivacy}. 
\provprivacy*

\begin{proof}
With loss of generality, we only prove the privacy guarantees for the device $C_i$. 
For ease of description, we set $\mathbf{r}^i=\mathbf{r}^i_*$ and $H^i = H^i_*$. 
Let $s^i$ be an indicator that takes value 1 if and only if 
$\mathcal{A}(\mathbf{r}^i) \neq u^i$, and 0 otherwise, i.e., $s^i = 1[\mathcal{A}(\mathbf{r}^i) \neq u^i]$. Now consider the joint entropy $H(\mathcal{A}(\mathbf{r}^i), u^i, s^i)$ of $\mathcal{A}(\mathbf{r}^i)$, $u^i$, and $s^i$. By decomposing it, we have 
\begin{align}
\label{eqn:entdecompo}
& H(s^i, u^i|\mathcal{A}(\mathbf{r}^i)) = H(u^i|\mathcal{A}(\mathbf{r}^i)) + H(s^i | u^i, \mathcal{A}(\mathbf{r}^i)) \notag \\ 
& =  H(s^i | \mathcal{A}(\mathbf{r}^i)) + H(u^i| s^i, \mathcal{A}(\mathbf{r}^i)),
\end{align}

Note that $H(s^i | u^i, \mathcal{A}(\mathbf{r}^i))=0$ as when $u^i$ and $\mathcal{A}(\mathbf{r}^i)$ are known, $S_i$ is also known. 
Similarly, 
\begin{align}
& H(u^i| s^i, \mathcal{A}(\mathbf{r}^i)) = Pr(s^i=1) H(u^i| s^i=1, \mathcal{A}(\mathbf{r}^i)) \notag \\
& \qquad + Pr(s^i=0) H(u^i| s^i=0, \mathcal{A}(\mathbf{r}^i)) = 0+0 = 0,
\end{align}
as knowing $s^i$ and $\mathcal{A}(\mathbf{r}^i)$, means we also know $u^i$.

Thus, Equation~\ref{eqn:entdecompo} reduces to $H(u^i|\mathcal{A}(\mathbf{r}^i))=  H(s^i | \mathcal{A}(\mathbf{r}^i))$. 
As conditioning does not increase entropy, i.e.,  
$H(s^i | \mathcal{A}(\mathbf{r}^i)) \leq H(s^i)$, we further have
\begin{align}
\label{eqn:finalone}
H(u^i|\mathcal{A}(\mathbf{r}^i)) \leq H(s^i).
\end{align}

On the other hand, using mutual information and entropy properties, we have 
$I(u^i; \mathcal{A}(\mathbf{r}^i)) = H(u^i) - H(u^i|\mathcal{A}(\mathbf{r}^i))$ and
$I(u^i; \mathbf{r}^i) = H(u^i) - H(u^i|\mathbf{r}^i)$. Hence, 
\begin{align}
\label{eqn:mi_en}
    I(u^i; \mathcal{A}(\mathbf{r}^i)) + H(u^i|\mathcal{A}(\mathbf{r}^i)) = I(u^i; \mathbf{r}^i) + H(u^i|\mathbf{r}^i).
\end{align}

Notice $\mathcal{A}(\mathbf{r}^i)$ is a random variable such that $u_i \perp \mathcal{A}(\mathbf{r}^i) | {\bf r}^i$. Hence, we have the Markov chain $u_i \rightarrow {\bf r}^i \rightarrow \mathcal{A}(\mathbf{r}^i)$.
Based on the data processing inequality in Lemma~\ref{lem:datainq}, we know $I(u^i; \mathcal{A}(\mathbf{r}^i)) \leq I(u^i; \mathbf{r}^i)$. 
Combining with Equation~\ref{eqn:mi_en}, we have 
\begin{align}
\label{eqn:finaltwo}
H(u^i|\mathcal{A}(\mathbf{r}^i)) \geq H(u^i|\mathbf{r}^i) = H^i.
\end{align}

Combing Equations~\ref{eqn:finalone} and~\ref{eqn:finaltwo}, we have $H(s^i) = H_2(Pr(s^i=1) )\geq H(u^i|\mathbf{r}^i)$,
which implies 
\begin{align}
    \label{eqn:final}
    & Pr(\mathcal{A}(\mathbf{r}^i) \neq u^i) = Pr(s^i=1) \notag \\ 
    & \geq H_2^{-1} (H(u^i|\mathbf{r}^i)) = H_2^{-1}(H^i),
\end{align}
where $H_2 (t) = -t \log_2 t - (1-t) \log_2 (1-t)$. 

Finally, by applying Lemma~\ref{lem:inveren}, we have 
$$Pr(\mathcal{A}(\mathbf{r}^i) \neq u^i) \geq \frac{H^i}{2 \log_2 (\frac{6}{H^i})}.$$
Hence the attribute privacy leakage is bounded by $Pr(\mathcal{A}(\mathbf{r}^i) = u^i) \leq 1 - \frac{H^i}{2 \log_2 ({6}/{H^i})}.$

\end{proof}

\begin{table}[!t]
\caption{Network architectures for the used datasets}
  \centering
    \begin{tabular}[t]{p{2cm}|p{1.8cm}|p{1.8cm}}
      \hline
    \centering\textbf{Feature Extractor} &\centering\textbf{Privacy Protection Network} &\textbf{Utility Preservation Network}\\
    \hline
    \hline
    \multicolumn{3}{c}{CIFAR-10}\\
    \hline
    \hline
    2xconv3-64 & 3xconv3-256 & conv3-16\\MaxPool & MaxPool & MaxPool\\
    \cline{1-3}
    2xconv3-128 & 3xconv3-512 & conv3-32\\
    MaxPool & MaxPool & MaxPool\\
    \cline{1-3}
    & 3xconv3-512 & 2xconv3-128\\
    & MaxPool & MaxPool\\
    \cline{2-3}
    & 2xlinear-4096 & 3xconv3-256\\
    &  & MaxPool\\
    \cline{2-3}
    & linear-$\#$priv. attri. values & 3xconv3-512\\
    &  & MaxPool\\
    \cline{2-3}
    &  & 3xconv3-512\\
    &  & MaxPool\\
    \cline{3-3}
    &  & linear-4096\\
    \cline{3-3}
    &  & linear-512\\
    \cline{3-3}
    &  &linear-1\\
    \cline{1-3}
    \hline
    \hline
    \multicolumn{3}{c}{Loans and Adult Income}\\
    \hline
    \hline
    linear-64  & linear-64 & linear-16\\
    \cline{1-3}
    linear-128 & linear-128 & linear-32\\
    \cline{1-3}
    & linear-4 & 2xlinear-128\\
    \cline{2-3}
    & linear-$\#$priv. attri. values& 3xlinear-256\\
    \cline{2-3}
    &  & 6xlinear-512\\
    \cline{3-3}
    &  & linear-4096\\
    \cline{3-3}
    &  & linear-512\\
    \cline{3-3}
    &  &linear-1\\
    \cline{1-3}
   \end{tabular}
   
   \label{tab:arch}
\end{table}

\section{Datasets and network architectures}
\label{appendix:Datasets}

\subsection{Detailed dataset descriptions}

\textbf{CIFAR-10 dataset~\cite{Krizhevsky09learningmultiple}.}
The CIFAR-10 (Canadian Institute For Advanced Research) dataset  contains 60,000 colored images of 32x32 resolution, which is split into
the training set with 50,000 images, and the testing set with 10,000 images. 
It is obtained from the \textit{torchvision.datasets} module, which provides a wide variety of built-in datasets. 
The dataset consists of images belonging to 10 classes: airplane, automobile, bird, cat, deer, dog, frog, horse, ship and truck. There are 6,000 images per class.  

For this dataset, the primary FL task has been established in accurately predicting the label of the image. The attribute to protect has been generated by the author, creating a binary attribute that is 1 if the image belongs to an animal and 0 otherwise.

\textbf{Loans dataset~\cite{10.5555/3157382.3157469}.}
This dataset is originally extracted from the loan-level Public Use Databases. The Federal Housing Finance Agency publishes these databases yearly, containing information about the Enterprises’ single family and multifamily mortgage acquisitions.  Specifically, the database used in this project is a single-family dataset and has a variety of features related to the person asking for a mortgage loan.  All the attributes in the dataset are numerical, so no preprocessing from this side was required. On the other hand, in order to create a balanced classification problem, some of the features were modified to have a similar number of observations belonging to all classes. We use 80\% data for training and 20\% for testing. 

The utility under this scope was measured in the system accurately predicting the affordability category of the person asking for a loan. This attribute is named \textit{Affordability}, and has three possible values: 0 if the person belongs to a mid-income family and asking for a loan in a low-income area, 1 if the person belongs to a low-income family and asking for a loan in a low-income area, and 2 if the person belongs to a low-income family and is asking for a loan not in a low-income area. The private attribute was set to be binary  \textit{Race}, being  White (0) or Not White (1).

\textbf{Adult Income dataset~\cite{Dua:2019}.}
This is a well-known dataset available in the UCI Machine Learning Repository. The dataset contains 32,561 observations each with 15 features, some of them numerical, other strings. Those attributes are not numerical were converted into categorical using an encoder. Again, we use the 80\%-20\% train-test split. 

The primary classification task is predicting if a person has an income above \$50,000, labeled as 1, or below, which is labeled as 0. The private attributes to predict are the \textit{Gender}, which is binary, and the \textit{Marital Status}, which has seven possible labels: 0 if Divorced, 1 if AF-spouse, 2 if Civil-spouse, 3 if Spouse absent, 4 if Never married, 5 if Separated, and 6 if Widowed. 

\subsection{Network architectures} 

The used network architectures for the three neural networks are in Table~\ref{tab:arch}.


\end{document}